\documentclass[aps,pre,groupedaddress,twocolumn,10pt]{revtex4-1}

\usepackage{amsmath}
\usepackage{amssymb}
\usepackage{amsthm}
\usepackage{bbm}
\usepackage{graphicx}

\usepackage{hyperref}
\usepackage[usenames,dvipsnames]{xcolor}
\hypersetup{colorlinks=true, linkcolor=BrickRed, urlcolor=blue!50!black, citecolor=blue!50!black}

\usepackage[normalem]{ulem}

\newcommand\diff{\mathrm{d}}
\renewcommand{\vec}[1]{\mathbf{#1}}
\renewcommand{\imath}[0]{\mathsf{i}}

\begin{document}

\title{Exact solution for the force-extension relation of a semiflexible polymer under compression}
\author{Christina Kurzthaler$^1$ and Thomas Franosch$^1$}
\affiliation{$^1$Institut f\"ur Theoretische Physik, Universit\"at Innsbruck, Technikerstra{\ss}e 21A, A-6020 Innsbruck, Austria}
\date{\today}

\begin{abstract}
Exact solutions for the elastic and thermodynamic properties for the wormlike chain model are elaborated in terms of Mathieu functions. 
The smearing of the classical Euler buckling instability for clamped polymers is analyzed for the force-extension relation. Interestingly, 
at strong compression forces the thermal fluctuations lead to larger elongations than for the elastic rod. The susceptibility defined as the 
derivative of the force-extension relation displays a prominent maximum at a force that approaches the critical Euler buckling force as the
 persistence length is increased. We also evaluate the excess entropy and heat capacity induced by the compresssion and find that they 
vary non-monotonically with the load. 
These findings are corroborated by pseudo-Brownian simulations. 
\end{abstract}
\maketitle

\section{Introduction}
The elasticity of biopolymers such as actin, microtubuli, or intermediate
filaments is responsible for the mechanical and structural stability of cells,
their motility and intracellular transport processes
\cite{Sackmann:1994,Brangwynne:2006, Bausch:2006,Fletcher:2010,Lieleg:2010,Nolting:2014}.  Networks of
these cytoskeletal polymers connected by regulatory proteins are exposed to mechanical
stresses, thereby exhibiting striking nonlinear elastic
behavior~\cite{MacKintosh:1995,Kroy:1996,Storm:2005,Claessens:2006,Chaudhuri:2007,Huisman:2008,Carillo:2013,Razbin:2015,Amuasi:2015,Plagge:2016}.  An
adequate characterization of the mechanical properties of these biological
assemblies constitutes a fundamental step toward the future design and synthesis
of artificial biomimetic materials~\cite{Ratner:2004,Ratner:2004:review}. Since
the macroscopic behavior of these filamentous materials strongly depends on the
properties of their constituents, a profound knowledge on the nature of
single filaments is necessary to fully understand the elasticity of these
networks~\cite{Hugel:2001}.

Due to their semiflexibility, biopolymers show a peculiar response to external
forces~\cite{Magnasco:1993,MacKintosh:2014, Razbin:2016}, where the static and dynamic
properties are dominated by their enthalpic elasticity, but conformational
entropy still plays a significant role. Experimental techniques such as
optical~\cite{Ashkin:1997,Mehta:1999} and magnetic tweezers~\cite{Gosse:2002},
transmission electron microscopy~\cite{Kuzumaki:2006}, and (atomic) force
spectroscopy~\cite{Janshoff:2000,Hugel:2001} have been used to quantify the
mechanical properties in terms of force-extension relations~\cite{Saleh:2015}.
These capture the  nonlinear effects emerging in the stretching and buckling of
semiflexible polymers such as
DNA~\cite{Marko:1995,Bouchiat:1999,Bustamante:2000}, actin~\cite{Liu:2002},
and synthetic carbon nanotubes~\cite{Kuzumaki:2006}, but also single molecules
such as titin~\cite{Kellermayer:1997}, which is an important component in
striated muscle tissues, and collagen~\cite{Sun:2002}, present in, e.g., skin and
bones. 

To characterize theoretically the physics of the nonlinear elastic behavior of
semiflexible polymers, the wormlike chain (WLC), also referred to as Kratky-Porod
model~\cite{Kratky:1949} has been analyzed in terms of the end-to-end
distribution function in the weakly-bending approximation ~\cite{Wilhelm:1996} as well as 
in terms of the exact low-order moments~\cite{,Hamprecht:2004, Spakowitz:2004,Spakowitz:2005,Mehraeen:2008}.
The WLC model has been shown to be a reliable model for semiflexible polymers,
as it reproduces, e.g., the mechanical behavior of DNA~\cite{Marko:1995} and
actin~\cite{Liu:2002}, while for microtubuli internal shearing of adjacent 
protofilaments leads to an apparent length dependence of the elastic 
moduli~\cite{Taute:2008,Pampaloni:2006}. In these studies, approximate solutions of the
force-extension relation of stretched polymers have been compared to
experimental data in order to determine mechanical properties such as the
persistence length.  

The approximations have been complemented by exact expressions for stretched
polymers in the plane~\cite{Prasad:2005}, while force-extension relations of
polymers under a compressive load have so far only been analyzed within the
regime of stiff polymers~\cite{Baczynski:2007,Emanuel:2007,Lee:2007,Bedi:2015}.  Therefore, a
characterization of the buckling behavior for a broad range of semiflexible
polymers is needed to predict elastic properties within experimental
observations. 

Here, we provide an exact analytical solution for the force-extension relation
of a compressed semiflexible polymer within the framework of the WLC model and
compare the results for polymers with different rigidity to computer
simulations. Boundary
conditions imposed within the experimental setup play a crucial role for the
response of polymers to external forces, as has been discussed earlier
for stretched polymers~\cite{Ghosh:2007}. To quantify this
effect, we also compare the force-extension relations  with two
clamped ends, one free and one clamped end, and two free ends.  
Thermodynamic properties, such as the in principle measurable excess heat capacity of the
semiflexible polymers induced by the compression, are also discussed. 

\section{The wormlike chain model\label{sec:model}}
For the elastic properties of a semiflexible polymer, we rely on the WLC model \cite{Kratky:1949}, where the bending energy of
an inextensible semiflexible polymer is expressed by its squared curvature,
\begin{align}
  \mathcal{H}_0	&= \frac{\kappa}{2}\int_0^L\diff s \ \Bigl(\frac{\diff \vec{u}(s)}{\diff s}\Bigr)^2.
\end{align}
Here, $s$ is the arc length, $\vec{u}(s) = \diff \vec{r}(s)/ \diff s$ denotes the tangent
vector of the polymer along its contour $\vec{r}(s)$, $\kappa$ the
bending stiffness of the polymer, and $L$ its contour length. The corresponding partition sum
$Z_0(\vec{u}_L,L|\vec{u}_0,0)$ of the polymer with initial orientation
$\vec{u}_0$ and final orientation $\vec{u}_L$ is obtained as a sum of
Boltzmann weights over all possible configurations, 
\begin{align}
  Z_0(\vec{u}_L,L|\vec{u}_0,0)  &= \int _{\vec{u}(0)=\vec{u}_0}^{\vec{u}(L)=\vec{u}_L}\mathcal{D}[\vec{u}(s)]\exp(-\mathcal{H}_0/k_\text{B}T),
\end{align}
where the inextensibility constraint $|\vec{u}(s)|=1$ has to be fulfilled. 
In addition, if the polymer is subject to a constant external force $\vec{F}$,
the stretching energy, 
\begin{align}
  \mathcal{H}_\text{force} &= -\int_0^L \diff s \ \vec{F}\cdot \vec{u}(s),
\end{align}
has to be added to the bending energy. 
Here, the force acts along a fixed direction $\vec{e}$
with strength $F$ such that $\vec{F}=F\vec{e}$. 
Collecting terms, the full Hamiltonian of the system reads
\begin{align}
  \frac{\mathcal{H}}{k_\text{B} T}  
  &= \int_0^L \diff s \ \Bigl[\frac{\kappa}{2k_\text{B}T}  \Bigl(\frac{\diff \vec{u}(s)}{\diff s}\Bigr)^2 -f \vec{e}\cdot \vec{u}(s)\Bigr], \label{eq:hamiltonian}
\end{align}
where $f = F/k_\text{B}T$ is the reduced force with units of an inverse length. The
persistence length $\ell_\text{p}=\kappa/k_\text{B}T$ for 3D, respectively 
$\ell_\text{p}=2\kappa/k_\text{B}T$ for 2D, corresponds to the 
decay length of the tangent-tangent correlations of the polymer and permits a classification
of polymers into three categories:
polymers with $\ell_\text{p}/L \ll 1$ are referred to as flexible and therefore more coil-like structured, 
$\ell_\text{p}/L \sim 1$ are semiflexible, and
$\ell_\text{p}/L \gg 1$ are stiff, hence, rodlike~\cite{Doi:1986, MacKintosh:2014}.  
The partition sum $Z(\vec{u}_L,L|\vec{u}_0,0)$ of such a polymer subject to an
external force is given by the path integral over all weighted, accessible chain
configurations. In particular, except for the prefactor $Z_0\equiv
Z_0(\vec{u}_L,L|\vec{u}_0,0)$, the partition sum can be viewed as the
generating function of a semiflexible polymer:
\begin{align}
  Z(\vec{u}_L,L|\vec{u}_0,0)  &= \int _{\vec{u}(0)=\vec{u}_0}^{\vec{u}(L)=\vec{u}_L}\mathcal{D}[\vec{u}(s)]\exp(-\mathcal{H}/k_\text{B}T),\label{eq:Z}\\
                              &= Z_0  \left\langle \exp\left(\int_0^L \diff s f \vec{e}\cdot\vec{u}\right)\right\rangle. 
\end{align}
The corresponding thermodynamic potential is the Gibbs free energy defined by 
\begin{align} 
G(T,F) &= -k_\text{B}T\ln[Z(\vec{u}_L,L|\vec{u}_0,0)], \label{eq:gibbs}
\end{align}
with fundamental relation 
\begin{align}
 \diff G &= -S\diff T -\langle X \rangle \diff F, 
\end{align}
where $S=S(T,F)$ denotes the entropy of the polymer and $\langle X\rangle :=\int_0^L\diff s \ \langle
\vec{e}\cdot\vec{u}(s)\rangle$ is the mean end-to-end distance projected onto the direction of the force. 

Hence, the mean projected end-to-end distance is obtained by
\begin{align}
  \langle X \rangle   &= -\left(\frac{\partial G}{\partial F}\right)_T =-\frac{1}{k_\text{B}T}\left(\frac{\partial G}{\partial f}\right)_T.\label{eq:X}
\end{align}
In particular, the response of the polymer to the applied force is encoded in the
isothermal susceptibility, 
\begin{align}
  \chi &= \left(\frac{\partial \langle X \rangle}{\partial F}\right)_T=\frac{1}{k_\text{B}T}\left(\frac{\partial \langle X \rangle}{\partial f}\right)_T, 
\end{align} 
which characterizes the strength of the response with
respect to the applied force. Furthermore it serves as an indicator for the buckling
force of semiflexible polymers in analogy to the critical Euler buckling force
for stiff rods~\cite{Landau:1986}.  

We can also obtain the linear susceptibility $\chi$ in terms of the 
fluctuation-response theorem~\cite{Doi:1986} for the semiflexible polymer,
\begin{align}
\chi(T,F) &= \frac{1}{k_\text{B}T}\left\langle \left( X-\langle X\rangle\right)^2\right\rangle.\label{eq:linearresponse}
\end{align} 

In addition to these elastic properties,  we obtain the change 
in the entropy with respect to the force at constant temperature by using the Maxwell relation:
\begin{align}
 \left(\frac{\partial S}{\partial F}\right)_T	&=  \left(\frac{\partial \langle X\rangle}{\partial T}\right)_F.
\end{align}
Thus, the excess entropy $\Delta S(T,F) = S(T,F)-S(T,F=0)$ can then be evaluated by 
\begin{align}
\Delta S(T,F) & =  \int_0^F\diff F'\left(\frac{\partial \langle X\rangle}{\partial T}\right)_{F'}.
\end{align}
Similarly, the experimentally accessible excess heat capacity 
$\Delta C_F(T,F) = C_F(T,F)-C_F(T,F=0)$ is obtained by
\begin{align}
\Delta C_F(T,F) 	&= \int_0^F \diff F' \ T\left(\frac{\partial^2 \langle X\rangle}{\partial T^2}\right)_{F'}. 
\end{align}  

\subsection{Analytic solution}
To determine the elastic and thermodynamic properties, the
partition sum has to be computed by solving for the path integral [Eq.~\eqref{eq:Z}].
Therefore, we discretize the path integral in terms of the arc-length~\cite{Kleinert:2009} and 
find that the partition sum obeys an equation of the Schr\"odinger type, 
describing the evolution of the partition sum along the contour 
of the polymer, 
\begin{align}
  \partial_s Z(\vec{u},s|\vec{u}_0,0)  &=[f\vec{e}\cdot\vec{u}+ \frac{k_\text{B}T}{2\kappa} \Delta_{\vec{u}}]Z(\vec{u},s|\vec{u}_0,0),
\end{align}
where $\Delta_\vec{u}$ denotes the angular part of the Laplacian, and with initial 
condition 
\begin{align}
  Z(\vec{u},s=0|\vec{u}_0,0) &= \delta(\vec{u},\vec{u_0}),
\end{align}
such that the delta function $\delta(\cdot,\cdot)$ 
enforces both orientations to coincide. 

Here we restrict the discussion to a polymer confined to a plane. 
Consequently, the
inextensibility constraint of the orientation can be parametrized in terms of a polar
angle, $\vec{u}=(\cos\varphi,\sin\varphi)^T$,  
were the angle $\varphi$ is measured with respect to the direction
of the applied force.
Thus, the evolution of the partition sum along the arc length reads 
\begin{align}
  \partial_s Z(\varphi,s|\varphi_0,0)  &=\left[f\cos(\varphi)+ \frac{1}{\ell_\text{p}} \partial^2_{\varphi}\right]Z(\varphi,s|\varphi_0,0),\label{eq:2d}
\end{align}
subject to the initial condition 
\begin{align}
  Z(\varphi,s=0|\varphi_0,0) &= \delta(\varphi-\varphi_0\text{ mod }2\pi).
\end{align}
The Fokker-Planck equation [Eq.~\eqref{eq:2d}] is reminiscent of the Schr\"odinger equation of a quantum pendulum~\cite{Aldrovandi:1980} and 
can be solved analytically by separation of variables in terms of appropriate angular eigenfunctions, 
$\exp(-\lambda s)z(\varphi)$.
Inserting this ansatz into Eq.~\eqref{eq:2d}, we obtain
\begin{align}
  \left[\lambda+f\cos\varphi+\frac{1}{\ell_\text{p}}\frac{\diff^2}{\diff \varphi^2}\right]z(\varphi)&=0.
\end{align}
Here we focus on a polymer exposed to a compressive load, in particular, we
rewrite the reduced force as $f=-|f|$.  A change of variables $x=\varphi/2$ and
rearranging terms, leads to the equation 
\begin{align}
  \left[\frac{\diff^2}{\diff x^2}+\left(a-2q \cos(2x)\right)\right]z(x)=0,\label{eq:Mathieu}
\end{align}
which is known as the Mathieu equation \cite{NIST:online,NIST:print} 
with deformation parameter $q=2|f|\ell_\text{p}$ 
and eigenvalue $a=4\ell_\text{p}\lambda $. Thus, the general solution is expressed as a linear combination of $\pi$-periodic even and odd Mathieu functions 
$\text{ce}_{2n}(q,x)$ and $\text{se}_{2n+2}(q,x)$
with corresponding eigenvalues $a_{2n}\equiv a_{2n}(q)=4\ell_\text{p}\lambda_n$ and $b_{2n+2}\equiv b_{2n+2}(q)=4\ell_\text{p}\lambda_n$~\cite{NIST:online,NIST:print}, respectively. 
Note that in the case of a polymer under tension, where the sign of the force is positive, $f=|f|$, 
a different change of variables $x=\pi/2-\varphi/2$ has 
to be used in order to arrive at the Mathieu equation~\cite{Prasad:2005}. 

The Mathieu functions are essentially deformed sines and cosines: 
\begin{align}
  \text{ce}_{2n}(q,x)   &= \sum_{m=0}^\infty A_{2m}^{2n}(q)\cos (2mx),\\
  \text{se}_{2n+2}(q,x) &= \sum_{m=0}^\infty B_{2m+2}^{2n+2}(q)\sin\bigl((2m+2)x\bigr).
\end{align}
The coefficients $A_{2m}^{2n}(q)$ and
$B_{2m+2}^{2n+2}(q)$ are determined by recurrence relations, 
which result from inserting the Fourier series into
Eq.~\eqref{eq:Mathieu}:
\begin{align}
  a_{2n}A_0^{2n}-qA_2^{2n}  &=0,\label{eq:rec1}\\
  (a_{2n}-4)A_2^{2n}-q(2A_0^{2n}+A_4^{2n})  &=0,\\
(a_{2n}-4m^2)A_{2m}^{2n}-q(A_{2m-2}^{2n}+A_{2m+2}^{2n})&=0, \quad m\geq 2,\label{eq:rec2}
\end{align}
and similar relations hold for the coefficients $B_{2m+2}^{2n+2}(q)$~\cite{NIST:online,
NIST:print}. The Mathieu
functions constitute a complete, orthogonal and normalized set of eigenfunctions:
$\int_{0}^{2\pi}\diff x \ \text{ce}_{2n}(q,x)\text{ce}_{2m}(q,x)=\delta_{nm}\pi$ 
and similarly for $\text{se}_{2n+2}(q,x)$.  
Hence, the full solution of Eq.~\eqref{eq:2d} in terms of the eigenfunctions reads 
\begin{align}
  \lefteqn{Z(\varphi_L,L|\varphi_0,0) =} \notag\\
    &   \frac{1}{2\pi}\sum_{n=0}^{\infty}\bigl[\text{ce}_{2n}(q,\varphi_0/2)\text{ce}_{2n}(q,\varphi_L/2)e^{-a_{2n}(q)L/4\ell_\text{p}}\notag\\
    &   +\text{se}_{2n+2}(q,\varphi_0/2)\text{se}_{2n+2}(q,\varphi_L/2)e^{-b_{2n+2}(q)L/4\ell_\text{p}}\bigr] .\label{eq:SumZ}
\end{align}
Since the eigenvalues are ordered with increasing $n$,
$a_0(q)<b_2(q)<a_2(q)<b_4(q)<...$, the series expansion converges and can be
evaluated numerically, which is discussed in more detail in Appendix~\ref{sec:numerics}.

We impose boundary conditions, that reflect
different experimental setups accounting for clamped or free orientations at
the ends of the polymer.  The presented partition sum [Eq.~\eqref{eq:SumZ}] represents the case of a
clamped polymer with given initial and final orientation.  This can be
complemented by two or more scenarios.  First, we look at a half-clamped polymer,
where the initial orientation is clamped and the final one is free. The corresponding 
partition sum $Z(L|\varphi,0)$ is obtained via integration over all final angles, 
\begin{align}
  &Z(L|\varphi_0,0)  = \int_{0}^{2\pi} \diff \varphi_L Z(\varphi_L,L|\varphi_0,0),\notag\\
  &=  \sum_{n=0}^{\infty}A_0^{2n}(q)\text{ce}_{2n}(q,\varphi_0/2)\exp[-a_{2n}(q)L/4\ell_\text{p}],\label{eq:SumZ2}
\end{align}
where $A_0^{2n}(q)$ results from the integral over the even Mathieu functions
and the odd Mathieu functions do not contribute anymore. 

Further, integrating also over the initial angle the partition sum 
corresponding to a free polymer reduces to
\begin{align}
Z(L)  &=  2\pi\sum_{n=0}^{\infty}[A_0^{2n}(q)]^2\exp[-a_{2n}(q)L/4\ell_\text{p}].\label{eq:SumZ3}
\end{align}
Interestingly, the same result occurs for the semiflexible polymer under tension~\cite{Prasad:2005}, 
which reflects that the polymer is free to align with the external force.
To evaluate the partition sums in Eqs.~\eqref{eq:SumZ} and~\eqref{eq:SumZ2} numerically, we rely 
on an implementation of the Mathieu functions in a computer algebra system~\cite{Mathematica}. 
However, to determine the Fourier coefficients $A_0^{2n}(q)$ in Eqs.~\eqref{eq:SumZ2} and~\eqref{eq:SumZ3} 
it is more efficient to solve numerically the eigenvalue problem of the recurrence relations [Eq.~\eqref{eq:rec1}-\eqref{eq:rec2}]. 

\subsection{Linear response}

The linear response $\chi(T,F=0)$ of the semiflexible polymer to the compression force can also be obtained exactly by 
expanding the full solution in Eqs.~\eqref{eq:SumZ},~\eqref{eq:SumZ2}, and~\eqref{eq:SumZ3} 
to second order in the force.
For a clamped polymer we use 
the expansions of the Mathieu functions for $\varphi_0=\varphi_L=0$ 
and the corresponding eigenvalues up to second order in the 
deformation parameter $q$ from Refs.~\cite{NIST:online,NIST:print}:
\begin{align}
a_{0}(q) &= -\frac{q^2}{2}, \ a_2(q) = 4+\frac{5 q^2}{12},\label{eq:exp1}\\ 
a_{2n}(q) &= 4n^2+ \frac{q^2}{2(4n^2-1)},  \\
\text{ce}_0(q,0) &= \frac{1}{\sqrt{2}}\left(1-\frac{q}{2}-\frac{q^2}{32}\right),\\
\text{ce}_2(q,0) &= 1+\frac{q}{6}-\frac{73q^2}{1152}, \\
\text{ce}_{2n}(q,0) &= 1-\frac{q}{2-8 n^2}+\frac{(8 n^2+1)q^2}{32 \left(1-4 n^2\right)^2 \left(n^2-1\right)}.\label{eq:exp2}
\end{align}
Note that for $\varphi_0=\varphi_L=0$ the odd Mathieu functions do not contribute 
to the partition sum. 
Inserting these expansions into the full solution [Eq.~\eqref{eq:SumZ}]
one can derive both the projected mean end-to-end distance as well as the 
force-free susceptibility [Eq.~\eqref{eq:linearresponse}]. 
The explicit expressions are rather lengthy for arbitrary parameters $L$ and $\ell_\text{p}$ 
and will not be shown here, but we will discuss them for a clamped polymer in Sec.~\ref{sec:elasticity}, 
Fig.~\ref{plot:baz}. 

For the half-clamped polymer ($\varphi_0=0$) only the first three terms, $n=0,1,2,$ contribute to the 
order considered to the partition sum in Eq.~\eqref{eq:SumZ2}. Here, we use the 
expansions up to second order in $q$ of the Fourier coefficients $A_0^0(q) = (1-q^2/16)/\sqrt{2}$, 
$A_0^2(q)=q/4$ and $A_0^4(q)=q^2/96$ obtained from Refs.~\cite{NIST:online,NIST:print}, and
inserted these together with the expansions from Eqs.\eqref{eq:exp1}-\eqref{eq:exp2} into the full solution
[Eq.~\eqref{eq:SumZ2}]. Thus, the partition sum reads
\begin{align}
Z&(L|\varphi_0=0,0) = \pi +\pi\ell_\text{p}\frac{|F|}{k_\text{B}T} \left(e^{-L/\ell_\text{p}}-1\right)\notag\\
&+\frac{\pi}{24} \ell_\text{p}^2 \left(\frac{|F|}{k_\text{B}T}\right)^2 \Bigl(\frac{12 L}{\ell_\text{p}}-9+ 8 e^{-L/\ell_\text{p}}+e^{-4L/\ell_\text{p}}\Bigr)\notag\\
&+\mathcal{O}\left(|F|^3\right), \label{eq:expandZ2}
\end{align}
which results in a projected mean end-to-end distance 
\begin{align}
\langle X\rangle &= \left(\ell_\text{p}-\ell_\text{p} e^{-L/\ell_\text{p}}\right)
+\frac{\ell_\text{p}|F|}{12k_\text{B}T} \Bigl\{\ell_\text{p} \Bigl[e^{-4 L/\ell} \Bigl(12 e^{2 L/\ell}\notag\\
&  \ \ \ -32 e^{3 L/\ell}-1\Bigr)+21\Bigr]-12 L\Bigr\}+\mathcal{O}\left(|F|^2\right),
\end{align}
and linear susceptibility 
\begin{align}
\chi(T,F=0) &= \frac{\ell_\text{p}}{12k_\text{B}T}\Bigl[\ell_\text{p} \Bigl(e^{-4 L/\ell_\text{p}}-12 e^{-2 L/\ell_\text{p}}\notag\\
& \ \ \ +32 e^{-L/\ell_\text{p}}-21\Bigr)+12 L\Bigr].
\end{align}

For the  free polymer only the first two terms contribute and the partition sum in Eq.~\eqref{eq:SumZ3} 
is evaluated similar
to the partition sum of a half-clamped polymer [Eq.~\eqref{eq:expandZ2}]. It
reduces to 
\begin{align}
 Z(L) &= \pi ^2+\frac{\pi ^2}{2} \ell_\text{p}^2 \left(\frac{|F|}{k_\text{B}T}\right)^2 \left(e^{-L/\ell_\text{p}}+\frac{L}{\ell_\text{p}}-1\right)+\mathcal{O}\left(|F|^3\right).
\end{align}
Here the projected mean end-to-end distance vanishes for zero force after averaging over the initial 
and final angles, 
\begin{align}
\langle X \rangle &= 
 \frac{\ell_\text{p}|F|}{k_\text{B}T}\left[\ell_\text{p} \left(1-e^{-L/\ell_\text{p}}\right)-L\right]+\mathcal{O}\left(|F|^2\right),
\end{align}
and the linear susceptibility reads
\begin{align}
\chi(T,F=0) &= \frac{\ell_\text{p}}{k_\text{B}T} \left[\ell_\text{p} \left(e^{-L/\ell_\text{p}}-1\right)+L\right].
\end{align}

\subsection{Pseudodynamics}
To validate our analytical results, we also simulate the semiflexible polymer 
under compression. To generate a set of representative conformations of the polymer
in equilibrium, we rely on Brownian dynamics simulations that yield the canonical
ensemble as stationary state. Since we are not interested in the dynamic properties
here, features such as hydrodynamic interactions, anisotropic friction, and the overall
translational motion are ignored. Thus the purpose of the pseudodynamics is merely 
to reproduce the equilibrium properties.  

We first discretize equidistantly the contour $L$ of the polymer in terms of the positions of the
beads $\{\vec{R}_i\}_{i=0}^{N}$ and corresponding tangent vectors $\{\vec{u}_i\}_{i=0}^{N-1}$, where
$\vec{u}_i=(\vec{R}_{i+1}-\vec{R}_{i})N/L$ is of unit length, $|\vec{u}_i|=1$. 

The pseudodynamics for the orientation of the $i$th bead is prescribed by the Langevin equation
in the It$\bar{\text{o}}$ sense,
 \begin{align}
   \diff & \vec{u}_i(t)   = -\hat{D}_\text{rot}\vec{u}_i(t)\diff t + \hat{D}_\text{rot}\vec{u}_i^\perp(t)\Bigl(\frac{N}{L}\frac{\ell_\text{p}}{2}\vec{u}_i^\perp(t)\cdot[\vec{u}_{i-1}(t)\notag\\
   & +\vec{u}_{i+1}(t)]-\frac{L}{N}|f|\vec{u}^\perp_i(t)\cdot\vec{e}\Bigr)\diff t + \sqrt{2\hat{D}_\text{rot}} \vec{u}_i^\perp(t) \diff \omega_i(t),\label{eq:seg}
\end{align}
for $i=1,\dotsc,N-2$ and appropriately modified Langevin equations at 
the boundary, $i=0,N-1$. Here $\vec{u}_i^\perp(t)$ is the unit orientation rotated clockwise by an angle of $\pi/2$, 
i.e. $\vec{u}_i^\perp(t) \cdot \vec{u}_i(t)= 0$ and  $\det[\left( \vec{u}_i(t),\vec{u}_i^\perp(t)\right)]=1$. 
Furthermore $\omega_i(t)$ is a Gaussian white noise process with
zero mean and variance $\langle\omega_i(t)\omega_j(t')\rangle=\delta_{ij}\delta(t-t')$ for
$i,j=0,\dotsc,N-1$.  
The scaled rotational diffusion of the orientations $\hat{D}_\text{rot}$ 
sets only the time scale in our case. We show in Appendix~\ref{ref:appendix}
that the Langevin equations [Eq.~\eqref{eq:seg}] in fact reproduce 
the canonical ensemble of the semiflexible polymer problem.  
\begin{figure}[htp]
\includegraphics[width = \linewidth,keepaspectratio]{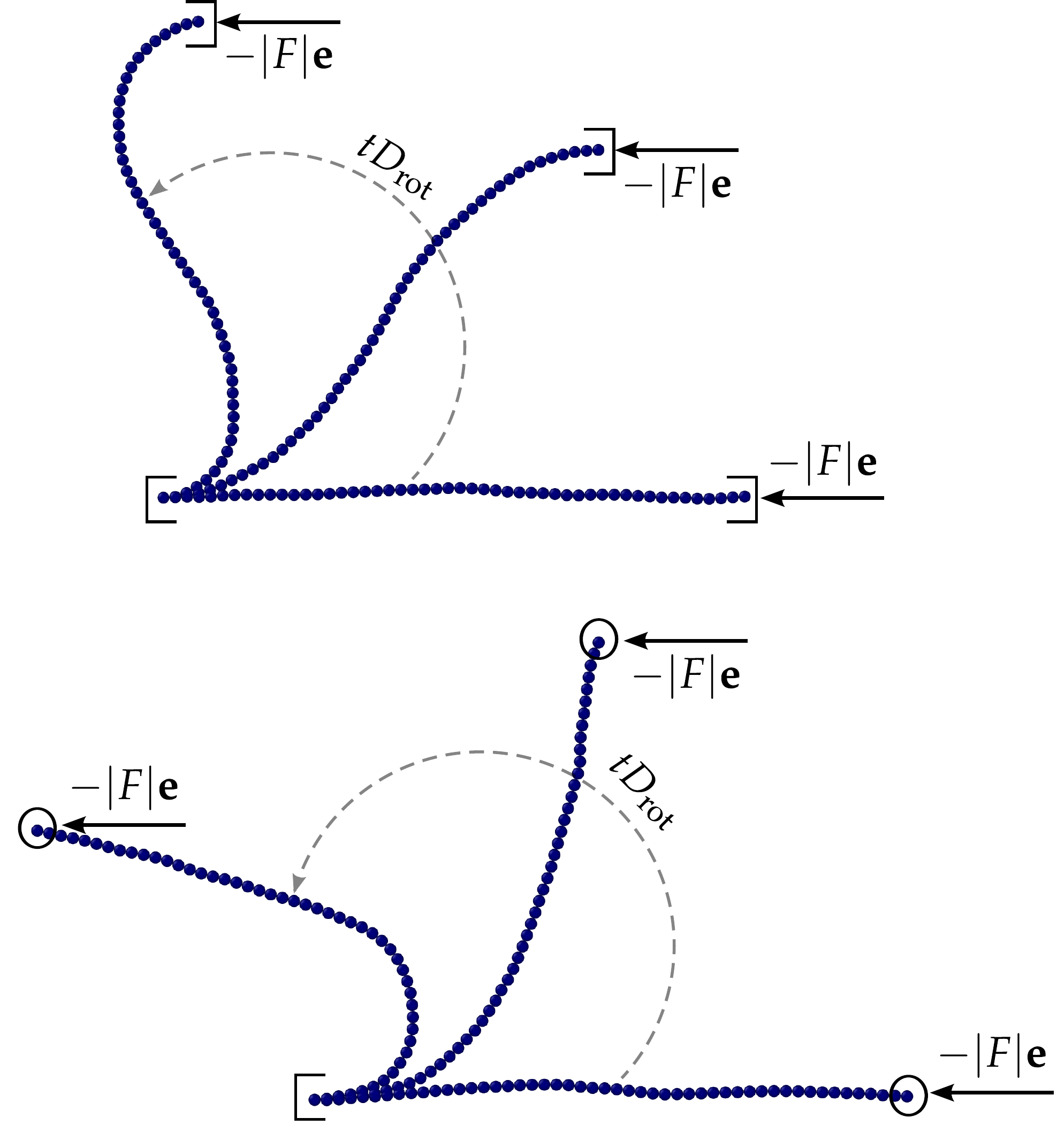}
\caption{Typical time evolution in simulation of a clamped (top) and a 
half-clamped (bottom) polymer starting from a straight configuration. 
The simulation is for a rather stiff polymer, $\ell_\text{p}=10L$, 
subject to a strong compression force $|F|= 2F_c$.}
\label{snapshot}
\end{figure}

The boundary conditions for a clamped polymer are set by fixed initial and
final orientation chosen into the direction antiparallel to the compression
force, in particular, $\vec{u}_0(t)=\vec{u}_{N-1}(t)=-\vec{e}$ for all times $t$.  
A half-clamped polymer fulfills the same boundary condition for the initial orientation,
$\vec{u}_0(t)=-\vec{e}$, whereas the final orientation can freely rotate. 

Simulation snapshots [Fig.~\ref{snapshot}] reveal 
that a semiflexible polymer shows qualitatively different buckling behavior
with respect to the boundary conditions.  Since
clamped polymers cannot rotate their ends, they exhibit an S-shaped configuration under
a compressive load. In contrast, if the second end of the polymer is free to rotate, it
can align with the direction of the applied force and a hook-shape configuration is
observed. Considering two free ends the polymer will just rotate under the force
and thus its behavior is expected to be identical to that of a polymer under tension. 

To obtain reliable statistics in equilibrated configurations 
we have conducted $120$ realizations of a polymer with $N=300$ segments and a time step of
$10^{-5}/\hat{D}_\text{rot}$ over a time horizon of $10^3/\hat{D}_\text{rot}$.
Measurements are taken after the polymer has reached equilibrium. For a free
polymer the simulations have been performed over a longer time horizon of $10^4/\hat{D}_\text{rot}$,
however, with fewer realizations, since the free rotation of the polymer 
requires long transients to reach equilibrium at low forces. 

In addition we have performed
Monte Carlo simulations, similar to Ref.~\cite{Baczynski:2007}, to crosscheck
the results. They confirm our analytical predictions for
clamped polymers, but they start to fail
in terms of convergence and statistics when refining the
discretization to capture the semiflexibility of the polymers and also the simulation of a free polymer has been
demanding.

\begin{figure*}[bt]
\includegraphics[width=\linewidth,keepaspectratio]{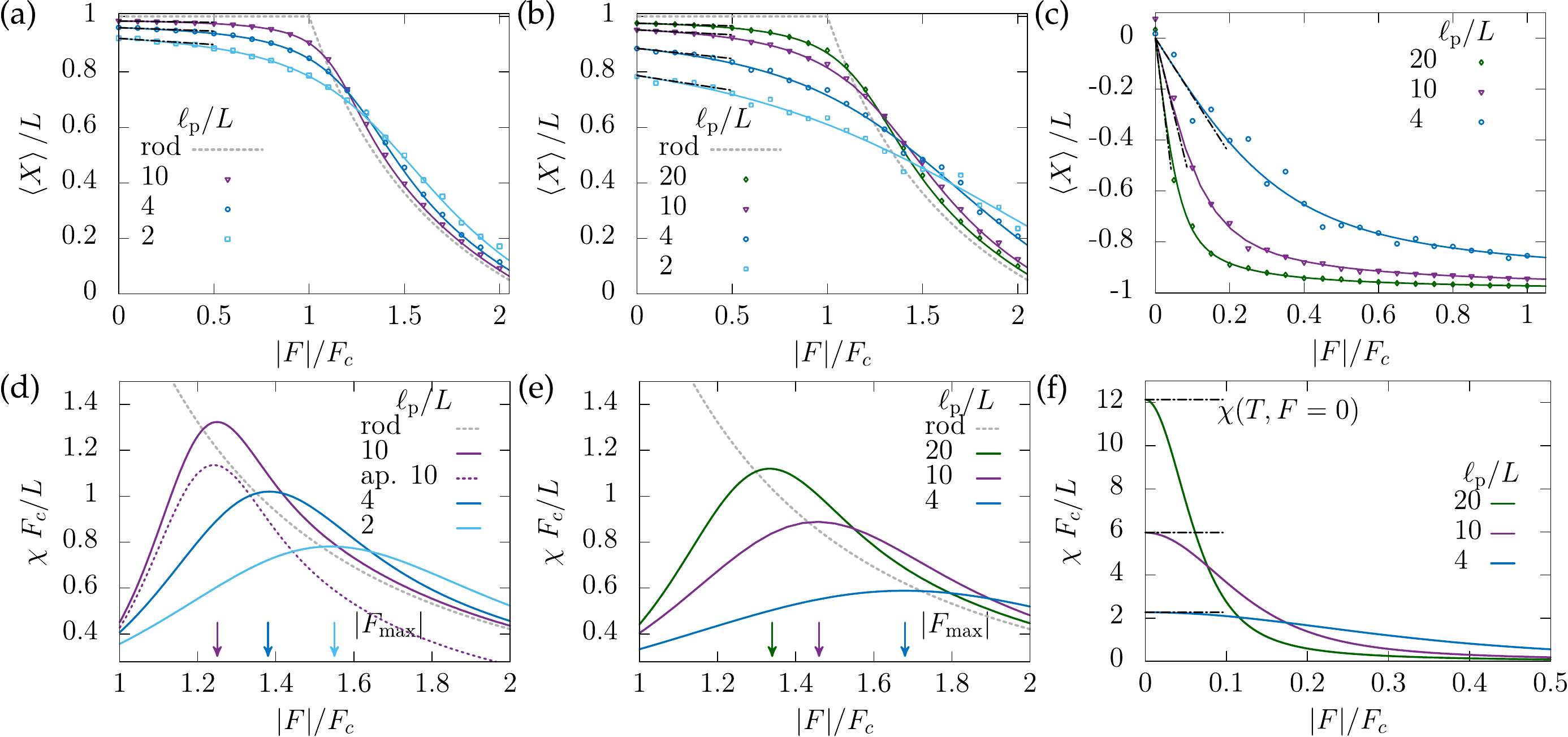} 
\caption{End-to-end distance in the direction of the applied force (top)  and
susceptibility (bottom) of polymers with different persistence lengths and
boundary conditions, clamped (a) and (d), half-clamped (b) and (e), and free ends (c) and (f).
The black dashed lines indicate the linear response and the gray dashed line corresponds to the classical Euler buckling, $F_c$ is the
classical Euler buckling force $F_c = \pi^2
\kappa/(\gamma L)^2$. In (c) and (f) we use $\gamma=2$ in $F_c$ to normalize the forces. 
The forces $|F_\text{max}|$ are extracted from
the maxima of the susceptibility.  In Fig. (d) 
the dashed line (ap. 10) represents the approximate susceptibility derived from 
the solution in Ref.~\cite{Baczynski:2007}.
Theory and pseudodynamics simulations are shown with lines and dots, respectively.\label{fig:all}} 
\end{figure*}

\subsection{Classical Euler buckling}
A stiff rod under compression at small forces does not yield at all but
starts to buckle at a critical force that is known as the Euler
buckling instability~\cite{Landau:1986}. The critical force is determined
by the rigidity $\kappa$ and the contour length $L$ by 
\begin{align}
F_c   &= \frac{\pi^2 \kappa}{(\gamma L)^2},
\end{align}
where $\gamma$ accounts for the different boundary conditions. 
In the case of a clamped rod, $\gamma = 1$, with one end fixed and one
translationally free; however, if one end of the rod is clamped and one
orientationally and translationally free, $\gamma = 2$ \cite{Landau:1986}.

For small temperatures the Fokker-Planck equation [Eq.~\eqref{eq:2d}] of the partition sum can be approximated for a stiff
rod by an Eikonal approximation~\cite{Kleinert:2009}. Here we set the partition sum $Z =\exp(-G/k_\text{B}T)$, 
neglect terms of higher order in the inverse temperature, and consequently obtain the approximate equation for the
Gibbs free energy,
\begin{align}
  \partial_s G-|F|\cos\varphi \ G+\frac{1}{2\kappa}\left(\partial_\varphi G\right)^2 &=0.
\end{align}
Using the method of characteristics leads to the equation of motion
for the angle,
\begin{align}
  \kappa \frac{\diff^2\varphi(s)}{\diff s^2}+|F|\sin\varphi(s)&=0,
\end{align} which is as anticipated reminiscent of the equation of motion of a classical
pendulum~\cite{Aldrovandi:1980}. The same equation results as Euler-Lagrange equation
minimizing the total energy [Eq.~\eqref{eq:hamiltonian}]. 
Starting from this equation, the force-extension relation for $|F|>F_c$ of a clamped and half-clamped rod
is well known~\cite{Landau:1986,Baczynski:2007},
\begin{align}
\langle X\rangle &= \sqrt{\frac{2\kappa}{|F|}}\int_0^{\varphi_{L/2}}\diff \varphi \ \frac{\cos\varphi}{\sqrt{\cos\varphi-\cos\varphi_{L/2}}}.
\end{align} 
These results will be compared to the elastic properties of
semiflexible polymers confined to two dimensions.

\section{Elastic properties\label{sec:elasticity}}
In this section we provide a discussion of the analytic solution and 
simulation results for the force-extension relation and the associated
susceptibility. 

\textit{Clamped polymer: }The mean projected end-to-end distance $\langle X \rangle$ of a
clamped polymer is obtained by numerical
differentiation of the Gibbs free energy [Eq.~\eqref{eq:X}].  
We find a smooth monotone crossover from
an almost stretched to a buckled configuration with increasing compression force, in
contrast to the classical Euler buckling instability, where the first
yielding takes place at the critical force $F_c$ [Fig.  \ref{fig:all} (a)]. 
Due to thermal fluctuations already for vanishing forces,
semiflexible polymers display undulations, hence, their projected end-to-end distance is shorter than their contour length,
$\langle X\rangle < L$. As anticipated, in the regime of small forces higher
flexibility leads to shorter mean end-to-end distances, which approach zero for completely flexible polymers, $\ell_\text{p}/L
\ll 1$. Qualitatively, the linear response behavior appears to be correct 
up to forces comparable to the classical Euler buckling force (see Figs.~\ref{fig:all} ~(a) and (b)).
Interestingly, for increasing forces, $|F| > F_c$, the end-to-end distance of
semiflexible polymers intersects with that of a stiff rod, indicating a
stiffening of the polymer due to thermal fluctuations. A similar trend occurs 
when comparing different persistence lengths, hence,
initially more flexible polymers yield more strongly while for large forces
they resist harder than stiffer polymers. 

\begin{figure}[t]
\includegraphics[width=\linewidth, keepaspectratio]{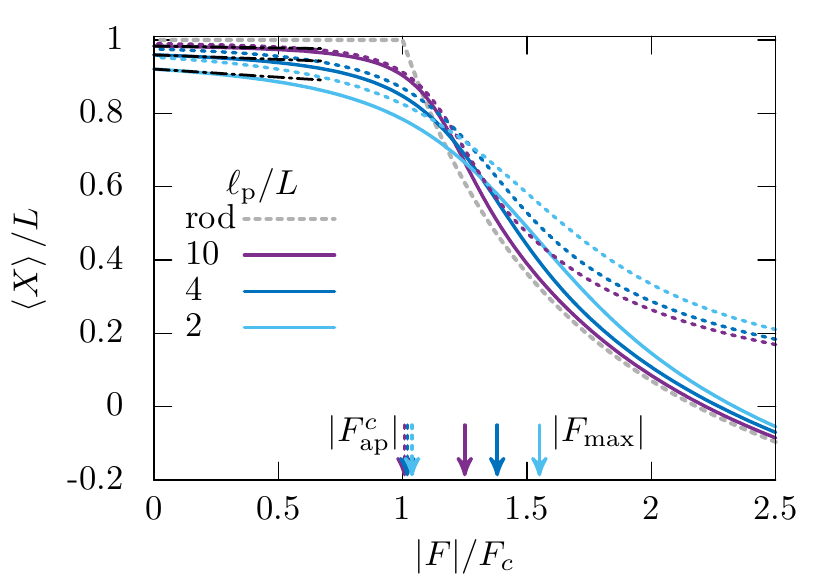}
\caption{Comparison of our analytic results to the approximate solution (dashed lines) 
of  Ref. \cite{Baczynski:2007} for different persistence lengths. $|F_\text{max}|$ 
denotes the force at maximal susceptibility from the analytic theory and $|F_\text{ap}^c|$ the 
critical force for the onset of buckling, taken from Ref.~\cite{Baczynski:2007}. The black dashed line
indicates the linear response, and the gray dashed line corresponds to the classical Euler buckling.  
\label{plot:baz}}
\end{figure}

Furthermore we compare the susceptibility $\chi$ of semiflexible polymers in
response to the compression force to the susceptibility of a stiff rod
[Fig.~\ref{fig:all} (d)]. The latter vanishes for forces smaller than the
critical Euler buckling force, reflecting that a classical rod does not 
yield. Directly at the transition the classical susceptibility assumes 
the value $\chi(F_c+0) = 2L/F_c$ and decreases monotonically for
larger forces.  
In contrast the susceptibility of semiflexible polymers
initially increases starting from $\chi(T,F=0)$ as predicted from the linear response (not shown), 
then displays a maximum at a force $|F_\text{max}|$, which exceeds the Euler buckling force,
$|F_\text{max}|>F_c$. 
Increasing the stiffness, this force 
approaches the critical buckling force $|F_\text{max}|\downarrow F_c$,
which suggests to use $|F_\text{max}|$ as a measure to charactzerize  
 the smooth transition for fluctuating semiflexible polymers. For simplicity,
we refer to $|F_\text{max}|$ as the buckling force also in the case of non-zero temperature. 
Interestingly, for stronger forces the projected end-to-end distance 
becomes even negative (see Fig.~\ref{plot:baz}), which reflects the S-shaped configuration in Fig.~\ref{snapshot}. 

\textit{Half-clamped polymer:}
Here, the force-extension relation behaves similarly to the one 
for clamped polymers (see Fig.~\ref{fig:all} (b)). Nevertheless half-clamped polymers
yield more strongly even after taking into account that
the classical buckling force is reduced by a factor of $\gamma^2=4$
due to the different boundary conditions. 
In fact, the data demonstrate that the semiflexibility becomes
more important for half-clamped polymers.   

Qualitatively, the monotonous transition from the straightened to the buckled
configuration of these polymers remains unchanged, intersections with the end-to-end
distance of a stiff rod persist also here. 
The corresponding susceptibilities [Fig.~\ref{fig:all} (e)] are less pronounced in comparison to 
clamped polymers of the same stiffness [Fig.~\ref{fig:all} (d)]. 
Also, here we characterize the transition by a buckling force $|F_\text{max}|$
taken as the maximum of the susceptibility.
For half-clamped ends the approach $|F_\text{max}|\downarrow F_c$ is slower
than for clamped ends. 

\textit{Free polymer:}
In contrast to clamped polymers, the force-extension relation of a free  polymer
exhibits qualitatively different behavior (see Fig.~\ref{fig:all} (c)). 
Direct inspection of the configurations (not shown) reveals that the polymer essentially 
rotates and aligns with the direction of the force. Thus, rather than compressing,
the force in fact stretches the polymer in the reverse direction, in particular, 
the projected end-to-end distances are all negative.   
The analytic solution for the case of pulling has been provided before~\cite{Prasad:2005},
however, only an approximate solution for rather flexible polymers of length
$L>4\ell_\text{p}$ has been discussed in more detail.
Thermal undulations prevent the polymer to be in a
completely straight configuration.  Stiffer polymers react to smaller forces
than more flexible polymers, and extend nearly to their full
contour length. For free ends the force-extension relations approach monotonically 
the case of a fully aligned rod as the stiffness increases, in contrast to the 
other two boundary conditions. Similarly, the susceptibility [Fig.~\ref{fig:all}~(f)], 
is always maximal at vanishing forces, where it assumes the value $\chi(T,F=0)$ as 
predicted by linear response theory, and is larger for increasing stiffness.
Correspondingly, no characteristic buckling 
force is extracted for this case, since after all the polymer stretches rather
than buckles in this case.

\textit{Approximate Solution:}
An approximate solution of the force-extension relation in the regime of rather
stiff polymers has been presented by Baczynski \textit{et al.}~\cite{Baczynski:2007} by integrating out short
wavelength fluctuations. Their results predict qualitatively
similar behavior for clamped polymers in terms of the smooth buckling
transition.  Yet, already in the regime for small forces $|F|\lesssim F_c$ 
deviations to our analytic theory become apparent (see Fig.~\ref{plot:baz}). The
differences decrease as stiffer polymers are considered. 
Their approximate solution for stiff polymers exhibits 
far weaker buckling than the predicted analytic solution and  
in addition the force-extension relation remains positive
for all forces.  This finding is not in agreement with the analytic solution 
for semiflexible polymers as well as with the
behavior of a classical stiff rod, which both predict a negative projected end-to-end distance for
large forces, $|F|\gg F_c$. Hence, the effect, that the polymer assumes and S-shaped configuration
in response to the applied force is not captured  within the approximate theory.

For both the analytic and the approximate solution, the force-extension curves
of a stiff rod and the semiflexible polymers intersect.  The corresponding
intersection force of the force-extension  has been computed in
Ref.~\cite{Baczynski:2007} and confirms that this intersection takes place
in the vicinity of the Euler-buckling instability. In contrast to the critical
buckling force $|F_\text{max}|$ defined by the maximal susceptibility, the
critical buckling force $|F_\text{ap}^c|$ of
Ref.~\cite{Baczynski:2007} has been defined as the onset of the buckling of the
polymer, shown in Fig.~\ref{plot:baz}.  Similar to our analysis, it exceeds the
critical Euler buckling force, but always remains smaller than the maximal buckling
force $|F_\text{ap}^c| < |F_\text{max}|$. Furthermore, the susceptibility extracted from 
the approximate theory deviates significantly from our analytic solution, yet,
the force at the maximal susceptibility remains similar (see
Fig.~\ref{fig:all}(d)).

\begin{figure}[h]
 \includegraphics[width=\linewidth, keepaspectratio]{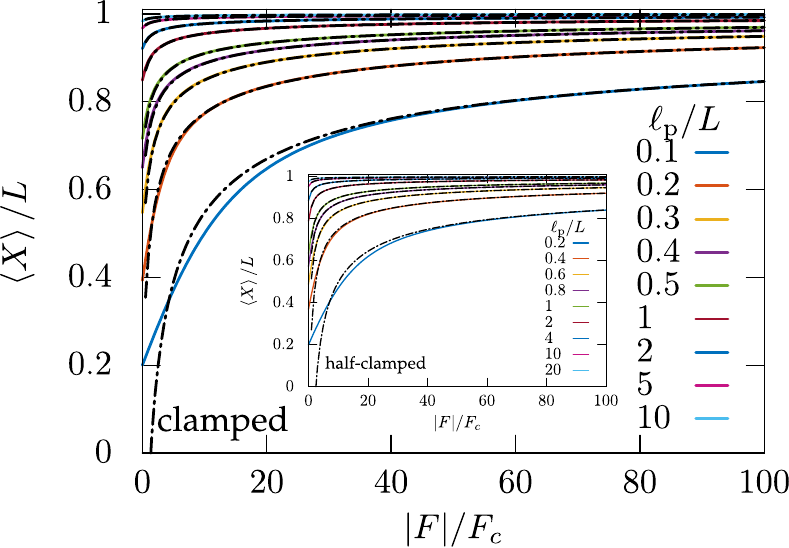}
\caption{End-to-end distance in the direction of the force for pulling of clamped and half-clamped (inset) polymers with different persistence lengths. 
We use the critical buckling force $F_c$ to normalize the forces. The black-dashed lines indicate the 
weakly-bending approximation.\label{plot:pull}}
\end{figure}

\textit{Pulled polymer: }We have also evaluated numerically for the first time the 
force-extension relation of a pulled polymer. The case of free ends 
under pulling is up to a sign identical to compression, as discussed above (see Fig.~\ref{fig:all}(c)).
In Fig.~\ref{plot:pull} we evaluate the analytic solution 
provided by Ref.~\cite{Prasad:2005} and compare it to the weakly-bending approximations~\cite{Smith:1992,Marko:1995} 
 for a clamped polymer, 
\begin{align}
\frac{\langle X\rangle}{L} &= 1+\frac{k_\text{B}T}{4L|F|}-\frac{\coth\left(L\sqrt{2|F|/\ell_\text{p}k_\text{B}T}\right)}{2\sqrt{2|F|\ell_\text{p}/k_\text{B}T}},
\end{align}
and a half-clamped polymer, 
\begin{align}
\frac{\langle X\rangle}{L} &= 1-\frac{\tanh\left(L\sqrt{2|F|/\ell_\text{p}k_\text{B}T}\right)}{2\sqrt{2|F|\ell_\text{p}/k_\text{B}T}};
\end{align}
see Appendix~\ref{app:weaklybending} for a derivation.
These approximations accurately reproduce the force-extension relation of pulled stiff polymers. 
However, the weakly-bending assumption is not fulfilled for more flexible polymers at small forces, which
display deviations from the exact solution (see Fig.~\ref{plot:pull}). 
The force-extension relation of a (half-) clamped polymer 
exhibits qualitatively similar behavior to that of a free polymer, 
and aligns along the direction of the applied force.

\section{Thermodynamic properties}
In addition to the elastic behavior, we discuss the thermodynamic
properties in terms of the excess entropy

\begin{align}
\Delta S(T,F) &= -\frac{\partial}{\partial T}\left[G(T,F)-G(T,F=0)\right],
\end{align}
and the excess heat capacity $\Delta C_F = T(\partial \Delta S/\partial T)_F$ 
in terms of our exact solution. 
\begin{figure}[h]
\includegraphics[width=\linewidth,keepaspectratio]{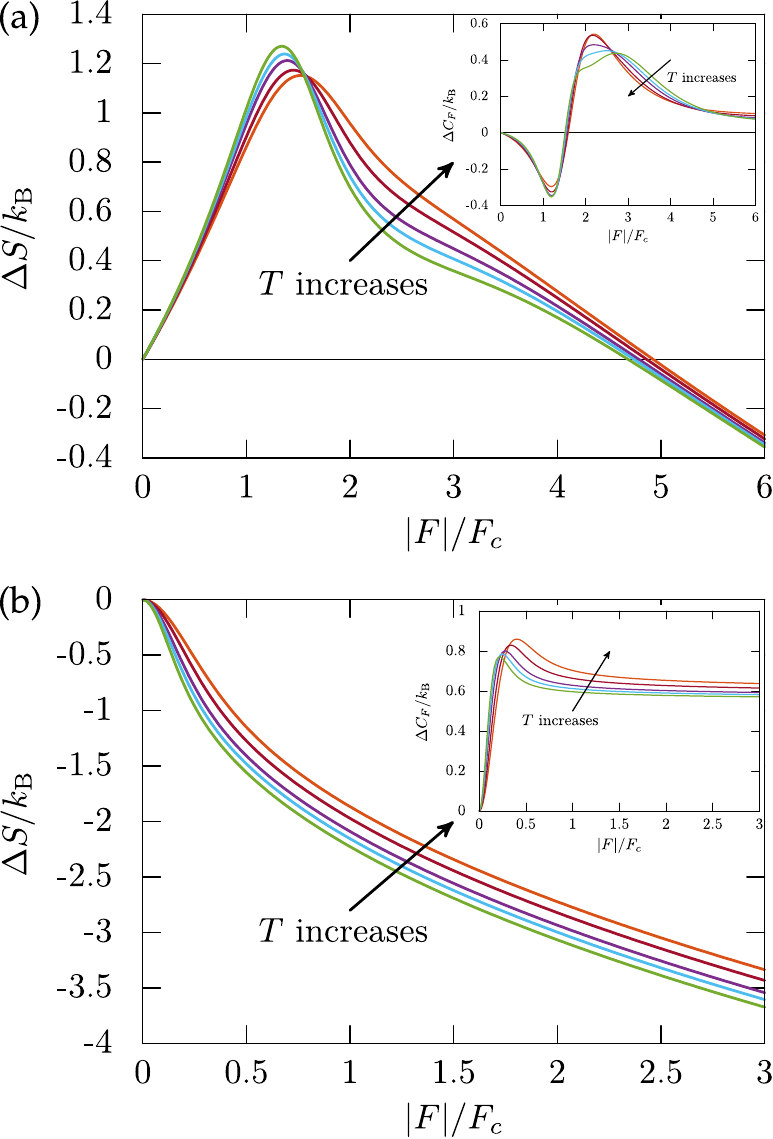}
\caption{Excess entropy $\Delta S$ and excess heat capacity $\Delta C_F$
(inset) for (a) clamped and (b) free polymers as a function of the force  $|F|/F_c$.
Temperature enters via $\ell_\text{p}/L=2\kappa/L k_\text{B}T= 1.4, \ 1.7, \ 2.0, \ 2.2, \ 2.5$.   
\label{fig:thermo}} 
\end{figure}

For clamped polymers, we observe an increase of the excess entropy $ \Delta S$ at small forces 
and a decrease at forces exceeding the Euler buckling force $|F|\gtrsim F_c $
(see Fig.~\ref{fig:thermo}~(a)). Thus, by applying a small force the number of accessible 
configurations of the polymer is increased, which reflects that by the S-shaped configurations
the fluctuations in the direction perpendicular to the force become more important.  
Yet, for strong forces the excess entropy decreases again and eventually even becomes negative.
In fact, the simulations show that for these strong forces the clamping suppresses the undulations
of the contour.

The temperature dependence of the excess entropy is 
encoded in the heat capacity $C_F(T,F)$. We find that 
the excess heat capacity is negative for forces 
$|F|\lesssim F_c$, whereas it becomes positive 
and displays a maximum for larger forces; see Fig.~\ref{fig:thermo}~(a)~(inset).
Thus, the initial increase of fluctuations upon compression 
is less important for more flexible polymers, while the suppression for large forces
is more relevant for stiffer polymers. 

Qualitatively similar results are obtained for half-clamped polymers
both for the excess entropy and the excess heat capacity (not shown). 

In contrast, free polymers behave qualitatively differently
in terms of their thermodynamic properties. In particular,
the excess entropy $\Delta S$ always decreases monotonically with increasing force (see
Fig.~\ref{fig:thermo}~(b)). Here the polymer reverses direction and is effectively 
under tension, such that the force suppresses the thermal undulations 
and the polymer straightens out.  The corresponding  excess heat capacity $\Delta C_F$ increases
and displays a prominent maximum before approaching a constant value for very large forces. 
As corroborated by the simulations this reflects that for more flexible polymers the 
thermal fluctuations are less suppressed than for stiffer ones (see
Fig.~\ref{fig:thermo}~(b)~(inset)).

\section{Summary and Conclusion} 
We have derived an analytic expression for the partition
sum of a semiflexible polymer under compression within the framework of
the WLC model for arbitrary stiffnesses.
The partition sum is represented as an expansion in terms of Mathieu functions,
which are the eigenfunctions of the associated Fokker-Planck equation. 
Since the eigenvalues form an ascending series to infinity, the infinite sum 
of decaying exponentials can be evaluated numerically (see Appendix~\ref{sec:numerics}). 
Elastic properties such as the force-extension relation, the susceptibility, and 
thermodynamic properties in terms of the excess entropy and excess heat capacity 
of the polymers with different persistence lengths, have been obtained as derivatives of the exact Gibbs free
energy.  These properties have been discussed for different boundary conditions
reflecting various experimental setups. 

Our results for the force-extension 
relation of (half-) clamped polymers predict a smooth crossover from an
equilibrated, almost straight rod to a buckled configuration.  Interestingly, 
the force-extension curves intersect in the vicinity of
the classical Euler buckling instability. 
Therefore, stiffer polymers resist more strongly for small forces, yet
at forces larger than the buckling force they yield more strongly
than flexible ones. 
Furthermore, the susceptibilities display a maximum in the vicinity 
of the classical Euler buckling force and by the fluctuation-response 
theorem this implies that the fluctuations of the projected mean end-to-end
distance become most important here.
Our analysis reveals that for large forces the weakly-bending approximation breaks down, 
as is apparent already for a classical elastic rod. 
Therefore the weakly-bending approximation remains valid 
only for large persistence lengths $\ell_\text{p} \gg L$ but 
is also restricted to small compression forces $|F|\lesssim F_c$. 

In contrast, a free polymer cannot resist a compression force, rather it rotates and 
aligns with the applied force. Therefore, our solution for this case 
reproduces the well-studied setup of a semiflexible polymer under tension.

Apart from the weakly-bending approximation~\cite{Baczynski:2007}, 
also different approximation schemes have been elaborated earlier to understand the 
behavior of rather stiff polymers close to the buckling transition. These
studies have incorporated corrections by thermal fluctuation to the theory of the 
classical Euler buckling instability~\cite{Lee:2007,Emanuel:2007}, which capture quantitatively 
the force-extension relation for forces much smaller and much 
larger than the critical Euler buckling force $F_c$
and compare it to simulations of semiflexible polymers modeled by a bead-spring chain~\cite{Emanuel:2007}. 
Including the lowest-order quartic mode
in the fluctuations predict qualitatively the same 
behavior as our analytical theory for planar inextensible polymers close to the critical Euler 
buckling instability~\cite{Lee:2007}.  
In addition, the buckling behavior of extensible polymers is also predicted to 
delay the buckling transition within the regime of small thermal 
fluctuations~\cite{Bedi:2015}. 
Alternatively to existing literature, one can also start from the probability distribution 
of a semiflexible polymer, which has been elaborated in the weakly-bending regime~\cite{Wilhelm:1996} and within a mean-field approach for filament 
inextensibility~\cite{Blundell:2009}, to compute approximately the force-extension relation.

However, these approximations~\cite{Wilhelm:1996,Baczynski:2007,Lee:2007,Emanuel:2007,Bedi:2015} 
remain valid for stiff polymers or away from the buckling transition only, 
as thermal fluctuations increase drastically for 
more flexible polymers at the maximum of the susceptibility.  In contrast, 
our analytical theory provides the force-extension relation
for the full range of semiflexible polymers.  

Also for pulled semiflexible polymers a lot of work has been done in elaborating the force-extension relation~\cite{Meng:2017}, ranging from weakly-bending approximations~\cite{Smith:1992,Marko:1995} to various approaches to account for the inextensibility of a wormlike chain~\cite{MacKintosh:1995,Ha:1997,Palmer:2008,Terentjev:2009}.
Here, we have evaluated for the first time the force-extension relation of pulled polymers using the 
analytical solution provided by Ref.~\cite{Prasad:2005}, where up to now only approximations of the fore-extension relations of rather flexible 
polymers have been discussed.
As we have shown in Appendix~\ref{sec:numerics}, for flexible polymers the numerical
evaluation of the partition sum reduces to a few terms, and can therefore be approximated due to the properties of the eigenvalues 
by the first term only, as has been observed earlier~\cite{Prasad:2005}. However, for increasing stiffness of the polymer  
more terms contribute to the partition sum, the approximation fails and the full solution is required to accurately 
reproduce these elastic properties. In the limiting case of stiff polymers, 
we find excellent agreement with the weakly-bending approximation for all forces (see Fig.~\ref{plot:pull}).

We anticipate, that our results serve as a reference for experimentally measured 
 force-extension relations, and allow determining the persistence length for a broad range of semiflexible
 polymers. We have provided tables (see Supplemental Material~\cite{[{See Supplemental Material at \url{http://link.aps.org/supplemental/10.1103/PhysRevE.95.052501} for tables of the force-extension relations}]supplement}) 
containing the numerical evaluation of 
 the force-extension relation for semiflexible polymers subject to compression as well as pulling forces, 
 in order to make the developed theory accessible and applicable to experimental observations.

Mathematically, the partition sum can be viewed as the Laplace transform of the 
distribution of the projected end-to-end distance. Therefore, the 
distribution can be obtained in principle by an inverse 
Laplace transform; however, this is numerically an ill-posed 
problem. Consequently, one should rather consider the characteristic function, i.e.,
the Fourier transform of the distribution,
which is obtained formally by analytic continuation in the force
to complex values. The Mathieu functions and the eigenvalues in this case
can still be obtained numerically by solving a matrix eigenvalue problem. 
This approach has been applied recently for the mathematical analog of 
a self-propelled particle~\cite{Kurzthaler:2016}.    

Our solution method is not restricted to the plane but can be 
extended to three dimensions. There the Mathieu functions of
the eigenvalue problem are replaced by generalized 
spheroidal wave functions~\cite{Leitmann:2016,Kurzthaler:2016}.   
Similarly, our approach can be extended in principle to account 
for a spontaneous curvature such that the classical reference 
system consists of a circular arc.

The equilibrium single-polymer behavior under tension or compression 
plays a crucial role as input for elastic properties of 
networks, and the regime of strong compression 
close to buckling may be useful to characterize the 
mechanical stability of such networks~\cite{MacKintosh:1995,Kroy:1996,Storm:2005,Claessens:2006,
Chaudhuri:2007,Huisman:2008,Carillo:2013,Razbin:2015,Amuasi:2015,Plagge:2016}.
In particular, networks are often modeled as entangled solutions of
single semiflexible polymers with entanglement points, where they cross or loop each other, branching points or
cross-links~\cite{MacKintosh:1995,Kroy:1996, Storm:2005, Huisman:2008,Carillo:2013,Razbin:2015,Amuasi:2015,Plagge:2016}. 
Already in equilibrium, single polymers experience a stretching or compression force induced 
by the surrounding network, and consequently the mean distance between, for example, two cross-links
differs from the contour length of the polymers. Here, our theoretical predictions for the elastic properties
of semiflexible polymers subject to different boundary conditions, which depend on the 
properties of the entanglement points or cross-links, allow adequate modeling and predict the equilibrium properties of polymer networks. 
Furthermore, exposing the network to mechanical stresses strongly depends on 
the force-extension relation of these individual filaments and crucially 
determines their elastic response. 
Up to now only approximate force-extension relations
have been applied in the analysis of networks~\cite{MacKintosh:1995,Kroy:1996, Storm:2005, Huisman:2008,Carillo:2013,Amuasi:2015}, 
whereas our characterization of (half-)clamped and free semiflexible polymers allows analyzing the 
elasticity of inhomogeneous networks composed of filaments 
with arbitrary stiffnesses. 
In particular, the clamped boundary conditions might serve as a starting point for modeling polymers 
connected by crosslinks, where their orientation is fixed due to the specific bonds. Differently, weaker entanglements can be accounted for by 
applying half-clamped boundary conditions, where one end of the polymer is allowed to rotate freely.
Hence, the theoretical predictions can be incorporated by 
using the tables provided in the Supplemental Material~\cite{supplement} 
in further modeling studies of networks, which would not alter the computational cost
of simulations significantly.
 
Similarly, the equilibrium behavior in free space constitutes the
reference for the relevant case of semiflexible polymers immersed 
in a dense crowded medium~\cite{Schobl:2014,Keshavarz:2016,Leitmann:2016}. 
Furthermore, the equilibrium properties should serve as the basis  also for dynamic
studies~\cite{Hallatschek:2007,Hallatschek:2007:2,Obermayer:2009}, in particular, 
the pseudodynamics we have derived here 
constitutes a convenient starting point to investigate 
the relaxation of undulation modes in the regime where the weakly-bending
approximation is no longer valid.

\begin{acknowledgements} 
We thank Sebastian Leitmann and Klaus Kroy for many discussions,
 and Sebastian Leitmann for a critical reading of the 
manuscript. This work has been supported by Deutsche Forschungsgemeinschaft
(DFG) via Contract No. FR1418/5-1.
\end{acknowledgements}

\appendix

\section{Numerical evaluation\label{sec:numerics}}
\begin{figure}[htp]
\centering
\includegraphics[width = \linewidth, keepaspectratio]{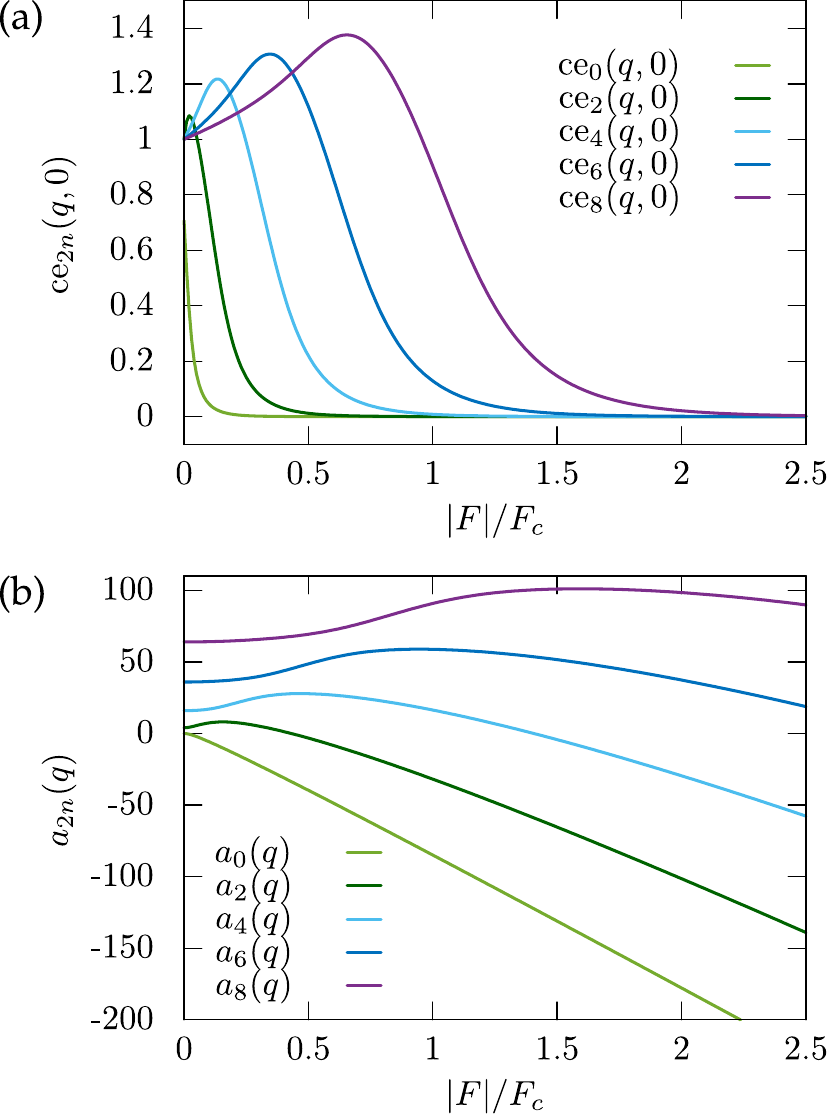}
\caption{(a) Even Mathieu functions $\text{ce}_{2n}(q,0)$ and (b) corresponding eigenvalues $a_{2n}(q)$ with deformation parameter 
$q=2|f|\ell_\text{p}$, reduced force $f=F/k_\text{B}T$, and persistence length $\ell_\text{p}= 10L$. 
Here, $F_c=\pi^2\kappa/L^2$ denotes the critical Euler buckling force of a clamped polymer.
\label{fig:mathieu}}
\end{figure}

Here, we discuss the numerical evaluation of the partition sums in Eqs.~\eqref{eq:SumZ}, \eqref{eq:SumZ2}, and \eqref{eq:SumZ3},
which are sums of relaxing exponentials with respect to the length $L$. Yet, the coefficients, i.e. the Mathieu functions 
($\text{ce}_{2n}(q,x)$ and $\text{se}_{2n+2}(q,x)$), as well as the exponents, i.e. the eigenvalues of the Mathieu functions ($a_{2n}(q)$ and $b_{2n+2}(q)$),
are non-monotonic functions with respect to the force $F$; see Fig.~\ref{fig:mathieu}. 

\begin{figure*}[bt]
\centering
\includegraphics[width = \linewidth, keepaspectratio]{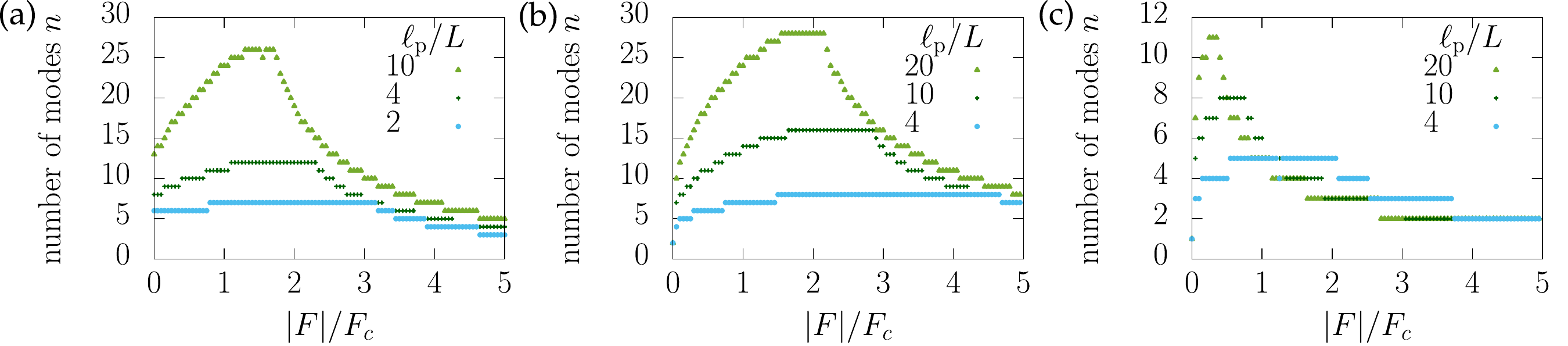}
\caption{Number of modes $n$ required to achieve an accuracy of
$|\langle X \rangle_{n+1}-\langle X \rangle_{n}|/L<10^{-8}$ with respect to the force $F$ 
for a clamped (a),  half-clamped (b), and  free polymer (c) of different persistence lengths $\ell_\text{p}$.
Here, $F_c=\pi^2\kappa/(\gamma L)^2$ denotes the critical Euler buckling force. In (c)
we use $\gamma = 2$ in $F_c$ to normalize the forces.\label{fig:acc}}
\end{figure*}

In particular, we find, that the zeroth eigenvalue, $a_0(q)$, remains negative for all forces, whereas the second, $a_2(q)$, and fourth eigenvalues, $a_4(q)$,
change sign within the parameter range considered. The other eigenvalues remain positive and increase with increasing mode $n$ for all relevant forces (see Fig.~\ref{fig:mathieu} (b)).
Consequently, the higher modes of the partition sum become exponentially suppressed, and therefore induce a natural cut-off of the infinite series [Eqs.~\eqref{eq:SumZ}, \eqref{eq:SumZ2}, and \eqref{eq:SumZ3}]. 

Moreover, with increasing mode $n$ the Mathieu functions become more important 
at intermediate and larger forces (see Fig.~\ref{fig:mathieu} (a)), and therefore many modes are 
expected to contribute to the partition sum. Interestingly, we observe that the 
number of modes necessary to achieve a desired accuracy of the elastic properties highly depends on the 
force. 

As an example, we consider the mean end-to-end distance $\langle X\rangle$, and determine the number of
modes necessary to obtain the accuracy $|\langle X \rangle_{n+1}-\langle X \rangle_{n}|/L<10^{-8}$, 
where $\langle X\rangle_n$ denotes the mean end-to-end distance with $n$ included modes.  
For a clamped and a half-clamped polymer, the number of modes increases and assumes a maximum in the vicinity of the critical Euler 
buckling force, and decreases again for larger forces; see Figs.~\ref{fig:acc}(a) and (b). Similarly, 
for a free polymer the number of modes also increases, yet at smaller forces, and decreases
for large forces [Fig.~\ref{fig:acc}(c)].

In addition, the number of modes rises drastically for stiffer polymers, which makes 
numerical evaluation more tedious. 
Furthermore, the number of modes for half-clamped boundary conditions is largest, whereas about half the modes
are required for a free polymer.

\section{Derivation of the pseudo dynamics\label{ref:appendix}}
In general, equations of motion in terms of stochastic differential equations can
be derived from the Fokker-Planck equation of the stochastic process
\cite{Gardiner:2009}. At this stage, only the equilibrium distribution of the
polymer subject to an external compression force is known and an equation of
motion for the probability density needs to be formulated.  Therefore, we
construct a Fokker-Planck equation starting with the requirement that the
known equilibrium distribution of the WLC model coincides with the stationary
distribution of the stochastic process. Therefore, we discretize the
polymer equidistantly in terms of the positions of the beads $\{\vec{R}_i\}_{i=0}^{N}$ and
corresponding tangent vectors $\{\vec{u}_i\}_{i=0}^{N-1}$, where $\vec{u}_i=(\vec{R}_{i+1}-\vec{R}_i)N/L$ 
with unit length $|\vec{u}_i|=1$. The discretized Hamiltonian takes the form 
\begin{align}
    \frac{\mathcal{H}(\{\vec{u}_i\}_{i=0}^{N-1})}{k_\text{B}T}&=
    \frac{\hat{\ell}_\text{p}}{4}\sum_{i=0}^{N-2}(\vec{u}_{i+1}-\vec{u}_{i})^2-\hat{f}\sum_{i=0}^{N-1}\vec{e}\cdot\vec{u}_i,\
\end{align}
where  $\hat{\ell}_\text{p}=\ell_\text{p}N/L$ and 
$\hat{f}=fL/N$ denote the scaled persistence length and 
force ($\hat{f}>0$ for tension and $\hat{f}<0$ for compression), respectively.

To fulfill the inextensibility constraint, we parametrize the orientation $\vec{u}_i =
(\cos\varphi_i,\sin\varphi_i)^T$ in terms of 
the polar coordinates $\{\varphi_i\}\equiv\{\varphi_i\}_{i=0}^{N-1}$
measured relative to the fixed direction $\vec{e}$ of the force. 
Therefore, the probability density in equilibrium is given by 
\begin{align}
            \mathbb{P}_\text{eq}(\{\varphi_i\}) &= Z^{-1} 
          \exp\left(-\frac{\mathcal{H}(\{\varphi_i\})}{k_\text{B}T}\right),
\end{align}
such that $\int \left[\prod_{i=0}^{N-1} \diff \varphi_i \right]  \mathbb{P}_\text{eq}(\{\varphi_i\})=1$. 

The discretized Hamiltonian in polar coordinates reads 
\begin{align}
            \frac{\mathcal{H}(\{\varphi_i\})}{k_\text{B}T}&=
            \frac{\hat{\ell}_p}{2}\sum_{i=0}^{N-2}[1-\cos(\varphi_{i+1}-\varphi_i)]
             -\hat{f}\sum_{i=0}^{N-1}\cos\varphi_i,
\end{align}
Next we derive the
Fokker-Planck equation describing the time evolution of the conditional
probability density
$\mathbb{P}\equiv\mathbb{P}(\{\varphi_i\},t|\{\varphi_i^0\})$ to find a
discretized polymer with orientations $\{\varphi_i\}$ at time $t$,
given that it was oriented with $\{\varphi_i^0\}$ at time $t=0$. It is
expressed by 
\begin{align}
  \partial_t \mathbb{P} = -\sum_{i=0}^{N-1}\partial_{\varphi_i} [U^i(\{\varphi_i\})\mathbb{P}],
\end{align}
where the velocity of the probability current is obtained by the friction law,
$U^i(\{\varphi_i\})=\sum_k K^{ik}(\{\varphi_i\})F_k(\{\varphi_i\})$. We choose for 
simplicity $\vec{K} = N\xi_r^{-1}\mathbb{I}$, which denotes the rotational mobility tensor
and $\vec{F}$ the forces,
\begin{align}
  F_k &=-\partial_{\varphi_k}\mathcal{H}-k_\text{B}T\partial_{\varphi_k}\ln\mathbb{P}.
\end{align}
Here, the first term corresponds to the mechanical forces while the second accounts for 
the thermal Brownian forces.
Thus one verifies that the equilibrium distribution $\mathbb{P}_\text{eq}$ corresponds to
the stationary distribution, $\partial_t\mathbb{P}_\text{eq}=0$.
Collecting results we find for the time evolution of the conditional probability density,
\begin{align}
\partial_t \mathbb{P}&= 
  \hat{D}_\text{rot}\sum_{i=0}^{N-1}\Bigl\{\partial_{\varphi_i}\Bigl[\frac{\hat{\ell}_\text{p}}{2}\left(\sin(\varphi_i-\varphi_{i-1})-\sin(\varphi_{i+1}-\varphi_i)\right)\notag\\
  & +\hat{f}\sin\varphi_i\Bigr]\mathbb{P}+\partial^2_{\varphi_i}\mathbb{P}\Bigr\},\label{eq:FPsim}
\end{align}
with scaled rotational diffusion coefficient
$\hat{D}_\text{rot}=Nk_\text{B}T/\xi_\text{r}$
and initial condition 
\begin{align}
  \mathbb{P}(\{\varphi_i\},t=0|\{\varphi_i^0\},0) &= \prod_{i=0}^{N-1}\delta(\varphi_i-\varphi_i^0\text{ mod }2\pi).
\end{align}
Starting from the Fokker-Planck equation [Eq.~\ref{eq:FPsim}] standard methods~\cite{Gardiner:2009} are used to obtain the Langevin
equations for the angles, governing the
pseudodynamics of a semiflexible, discretized polymer,
\vspace{-0.2cm}
\begin{align}
\diff&\varphi_i(t) = - \hat{D}_\text{rot}\Bigl[\frac{\hat{\ell}_\text{p}}{2}\left(\sin(\varphi_i-\varphi_{i-1}) -\sin(\varphi_{i+1}-\varphi_i)\right)\notag\\
&+\hat{f}\sin\varphi_i\Bigr]\diff t+\sqrt{2\hat{D}_\text{rot}}\diff\omega_i(t) ,
\end{align}
where $\diff \omega_i (t)$ is the increment of a Gaussian white noise process $\omega_i(t)$ with zero mean $\langle \omega_i(t)\rangle =0$ and
delta-correlated variance $\langle \omega_i(t) \omega_j(s)\rangle = \delta_{ij} \delta(t-s)$ for $i,j=0,\ldots, N-1$. These equations can be transformed by It$\bar{\text{o}}$'s lemma~\cite{Gardiner:2009}
to the equations for the tangent vectors in Eq.~\eqref{eq:seg}. 

\section{Force-extension relation for a pulled polymer in the weakly-bending regime\label{app:weaklybending}}
In the weakly-bending regime of a pulled polymer, $f=|f|$, we can approximate Eq.~\eqref{eq:2d} by 
\begin{align}
 \partial_s Z(\varphi, s|\varphi_0,0)	&= \left[\frac{1}{\ell_\text{p}}\partial^2_\varphi+|f|(1-\frac{1}{2}\varphi^2)\right]Z(\varphi, s|\varphi_0,0),
\end{align}
and use the Gaussian ansatz 
\begin{align}
Z(\varphi, s|\varphi_0,0) = \exp\left[-M(s)\varphi^2/2+\Gamma(s)\right].
\end{align}
Then the inverse variance $M(s)$ and the normalization $\Gamma(s)$ have to fulfill the 
equations of motion 
\begin{align}
\frac{\diff}{\diff s} M(s) &= |f|-2\frac{M(s)^2}{\ell_\text{p}},\\
 \frac{\diff}{\diff s}\Gamma(s) &= |f|-\frac{M(s)}{\ell_\text{p}}.
\end{align}
Using the initial condition $Z(\varphi, s=0|\varphi_0=0,0)=\delta(\varphi)$, the solutions of these differential 
equations read
\begin{align}
 M(s)		&= \sqrt{\frac{|f|\ell_\text{p}}{2}}\coth\left(\sqrt{\frac{2|f|}{\ell_\text{p}}}s\right),\\
\Gamma(s) 	&= |f|s-\frac{1}{2}\ln\sinh\left(\sqrt{\frac{2|f|}{\ell_\text{p}}}s\right)+\frac{1}{4}\ln\left(\frac{\ell_\text{p}|f|}{8\pi^2}\right).
\end{align}
To obtain the partition sum for a half-clamped polymer with contour 
length $L$, we average over the final orientation,
\begin{align}
 Z(L|\varphi_0,0)	
			&\approx \int_{-\infty}^\infty \diff\varphi_L\exp\left[-M(L)\varphi_L^2/2+\Gamma(L)\right]\notag\\
			&= \exp\left[\Gamma(L)\right]\sqrt{\frac{2\pi}{M(L)}},
\end{align}
which is correct if the angular fluctuations are small, $M(s)\gg 1$. This is fulfilled for stiff polymers  $L/\ell_\text{p}\ll 1$ 
or strong pulling $\ell_\text{p}|f|\gg1$.
Finally, we obtain the force-extension relations for (half-) clamped polymers in the weakly-bending regime
by taking the derivative of the Gibbs free energy [Eq.~\eqref{eq:gibbs}] with respect to the force (see Sec.~\ref{sec:model}). 


\begin{thebibliography}{70}%
\makeatletter
\providecommand \@ifxundefined [1]{%
 \@ifx{#1\undefined}
}%
\providecommand \@ifnum [1]{%
 \ifnum #1\expandafter \@firstoftwo
 \else \expandafter \@secondoftwo
 \fi
}%
\providecommand \@ifx [1]{%
 \ifx #1\expandafter \@firstoftwo
 \else \expandafter \@secondoftwo
 \fi
}%
\providecommand \natexlab [1]{#1}%
\providecommand \enquote  [1]{``#1''}%
\providecommand \bibnamefont  [1]{#1}%
\providecommand \bibfnamefont [1]{#1}%
\providecommand \citenamefont [1]{#1}%
\providecommand \href@noop [0]{\@secondoftwo}%
\providecommand \href [0]{\begingroup \@sanitize@url \@href}%
\providecommand \@href[1]{\@@startlink{#1}\@@href}%
\providecommand \@@href[1]{\endgroup#1\@@endlink}%
\providecommand \@sanitize@url [0]{\catcode `\\12\catcode `\$12\catcode
  `\&12\catcode `\#12\catcode `\^12\catcode `\_12\catcode `\%12\relax}%
\providecommand \@@startlink[1]{}%
\providecommand \@@endlink[0]{}%
\providecommand \url  [0]{\begingroup\@sanitize@url \@url }%
\providecommand \@url [1]{\endgroup\@href {#1}{\urlprefix }}%
\providecommand \urlprefix  [0]{URL }%
\providecommand \Eprint [0]{\href }%
\providecommand \doibase [0]{http://dx.doi.org/}%
\providecommand \selectlanguage [0]{\@gobble}%
\providecommand \bibinfo  [0]{\@secondoftwo}%
\providecommand \bibfield  [0]{\@secondoftwo}%
\providecommand \translation [1]{[#1]}%
\providecommand \BibitemOpen [0]{}%
\providecommand \bibitemStop [0]{}%
\providecommand \bibitemNoStop [0]{.\EOS\space}%
\providecommand \EOS [0]{\spacefactor3000\relax}%
\providecommand \BibitemShut  [1]{\csname bibitem#1\endcsname}%
\let\auto@bib@innerbib\@empty
\bibitem [{\citenamefont {Sackmann}(1994)}]{Sackmann:1994}%
  \BibitemOpen
  \bibfield  {author} {\bibinfo {author} {\bibfnamefont {E.}~\bibnamefont
  {Sackmann}},\ }\href {\doibase 10.1002/macp.1994.021950103} {\bibfield
  {journal} {\bibinfo  {journal} {Macromolecular Chemistry and Physics}\
  }\textbf {\bibinfo {volume} {195}},\ \bibinfo {pages} {7} (\bibinfo {year}
  {1994})}\BibitemShut {NoStop}%
\bibitem [{\citenamefont {Brangwynne}\ \emph {et~al.}(2006)\citenamefont
  {Brangwynne}, \citenamefont {MacKintosh}, \citenamefont {Kumar},
  \citenamefont {Geisse}, \citenamefont {Talbot}, \citenamefont {Mahadevan},
  \citenamefont {Parker}, \citenamefont {Ingber},\ and\ \citenamefont
  {Weitz}}]{Brangwynne:2006}%
  \BibitemOpen
  \bibfield  {author} {\bibinfo {author} {\bibfnamefont {C.~P.}\ \bibnamefont
  {Brangwynne}}, \bibinfo {author} {\bibfnamefont {F.~C.}\ \bibnamefont
  {MacKintosh}}, \bibinfo {author} {\bibfnamefont {S.}~\bibnamefont {Kumar}},
  \bibinfo {author} {\bibfnamefont {N.~A.}\ \bibnamefont {Geisse}}, \bibinfo
  {author} {\bibfnamefont {J.}~\bibnamefont {Talbot}}, \bibinfo {author}
  {\bibfnamefont {L.}~\bibnamefont {Mahadevan}}, \bibinfo {author}
  {\bibfnamefont {K.~K.}\ \bibnamefont {Parker}}, \bibinfo {author}
  {\bibfnamefont {D.~E.}\ \bibnamefont {Ingber}}, \ and\ \bibinfo {author}
  {\bibfnamefont {D.~A.}\ \bibnamefont {Weitz}},\ }\href {\doibase
  10.1083/jcb.200601060} {\bibfield  {journal} {\bibinfo  {journal} {The
  Journal of cell biology}\ }\textbf {\bibinfo {volume} {173}},\ \bibinfo
  {pages} {733—741} (\bibinfo {year} {2006})}\BibitemShut {NoStop}%
\bibitem [{\citenamefont {Bausch}\ and\ \citenamefont
  {Kroy}(2006)}]{Bausch:2006}%
  \BibitemOpen
  \bibfield  {author} {\bibinfo {author} {\bibfnamefont {A.}~\bibnamefont
  {Bausch}}\ and\ \bibinfo {author} {\bibfnamefont {K.}~\bibnamefont {Kroy}},\
  }\href {\doibase 10.1038/nphys260} {\bibfield  {journal} {\bibinfo  {journal}
  {Nature Physics}\ }\textbf {\bibinfo {volume} {2}},\ \bibinfo {pages} {231}
  (\bibinfo {year} {2006})}\BibitemShut {NoStop}%
\bibitem [{\citenamefont {Fletcher}\ and\ \citenamefont
  {Mullins}(2010)}]{Fletcher:2010}%
  \BibitemOpen
  \bibfield  {author} {\bibinfo {author} {\bibfnamefont {D.~A.}\ \bibnamefont
  {Fletcher}}\ and\ \bibinfo {author} {\bibfnamefont {R.~D.}\ \bibnamefont
  {Mullins}},\ }\href {\doibase 10.1038/nature08908} {\bibfield  {journal}
  {\bibinfo  {journal} {Nature}\ }\textbf {\bibinfo {volume} {463}},\ \bibinfo
  {pages} {485} (\bibinfo {year} {2010})}\BibitemShut {NoStop}%
\bibitem [{\citenamefont {Lieleg}\ \emph {et~al.}(2010)\citenamefont {Lieleg},
  \citenamefont {Claessens},\ and\ \citenamefont {Bausch}}]{Lieleg:2010}%
  \BibitemOpen
  \bibfield  {author} {\bibinfo {author} {\bibfnamefont {O.}~\bibnamefont
  {Lieleg}}, \bibinfo {author} {\bibfnamefont {M.~M. A.~E.}\ \bibnamefont
  {Claessens}}, \ and\ \bibinfo {author} {\bibfnamefont {A.~R.}\ \bibnamefont
  {Bausch}},\ }\href {\doibase 10.1039/B912163N} {\bibfield  {journal}
  {\bibinfo  {journal} {Soft Matter}\ }\textbf {\bibinfo {volume} {6}},\
  \bibinfo {pages} {218} (\bibinfo {year} {2010})}\BibitemShut {NoStop}%
\bibitem [{\citenamefont {Nolting}\ \emph {et~al.}(2014)\citenamefont
  {Nolting}, \citenamefont {M{\"o}bius},\ and\ \citenamefont
  {K{\"o}ster}}]{Nolting:2014}%
  \BibitemOpen
  \bibfield  {author} {\bibinfo {author} {\bibfnamefont {J.-F.}\ \bibnamefont
  {Nolting}}, \bibinfo {author} {\bibfnamefont {W.}~\bibnamefont {M{\"o}bius}},
  \ and\ \bibinfo {author} {\bibfnamefont {S.}~\bibnamefont {K{\"o}ster}},\
  }\href {\doibase 10.1016/j.bpj.2014.10.039} {\bibfield  {journal} {\bibinfo
  {journal} {Biophysical Journal}\ }\textbf {\bibinfo {volume} {107}},\
  \bibinfo {pages} {2693 } (\bibinfo {year} {2014})}\BibitemShut {NoStop}%
\bibitem [{\citenamefont {MacKintosh}\ \emph {et~al.}(1995)\citenamefont
  {MacKintosh}, \citenamefont {K\"as},\ and\ \citenamefont
  {Janmey}}]{MacKintosh:1995}%
  \BibitemOpen
  \bibfield  {author} {\bibinfo {author} {\bibfnamefont {F.~C.}\ \bibnamefont
  {MacKintosh}}, \bibinfo {author} {\bibfnamefont {J.}~\bibnamefont {K\"as}}, \
  and\ \bibinfo {author} {\bibfnamefont {P.~A.}\ \bibnamefont {Janmey}},\
  }\href {\doibase 10.1103/PhysRevLett.75.4425} {\bibfield  {journal} {\bibinfo
   {journal} {Phys. Rev. Lett.}\ }\textbf {\bibinfo {volume} {75}},\ \bibinfo
  {pages} {4425} (\bibinfo {year} {1995})}\BibitemShut {NoStop}%
\bibitem [{\citenamefont {Kroy}\ and\ \citenamefont {Frey}(1996)}]{Kroy:1996}%
  \BibitemOpen
  \bibfield  {author} {\bibinfo {author} {\bibfnamefont {K.}~\bibnamefont
  {Kroy}}\ and\ \bibinfo {author} {\bibfnamefont {E.}~\bibnamefont {Frey}},\
  }\href {\doibase 10.1103/PhysRevLett.77.306} {\bibfield  {journal} {\bibinfo
  {journal} {Phys. Rev. Lett.}\ }\textbf {\bibinfo {volume} {77}},\ \bibinfo
  {pages} {306} (\bibinfo {year} {1996})}\BibitemShut {NoStop}%
\bibitem [{\citenamefont {Storm}\ \emph {et~al.}(2005)\citenamefont {Storm},
  \citenamefont {Pastore}, \citenamefont {MacKintosh}, \citenamefont
  {Lubensky},\ and\ \citenamefont {Janmey}}]{Storm:2005}%
  \BibitemOpen
  \bibfield  {author} {\bibinfo {author} {\bibfnamefont {C.}~\bibnamefont
  {Storm}}, \bibinfo {author} {\bibfnamefont {J.~J.}\ \bibnamefont {Pastore}},
  \bibinfo {author} {\bibfnamefont {F.~C.}\ \bibnamefont {MacKintosh}},
  \bibinfo {author} {\bibfnamefont {T.~C.}\ \bibnamefont {Lubensky}}, \ and\
  \bibinfo {author} {\bibfnamefont {P.~A.}\ \bibnamefont {Janmey}},\ }\href
  {\doibase 10.1038/nature03521} {\bibfield  {journal} {\bibinfo  {journal}
  {Nature}\ }\textbf {\bibinfo {volume} {435}},\ \bibinfo {pages} {191}
  (\bibinfo {year} {2005})}\BibitemShut {NoStop}%
\bibitem [{\citenamefont {Claessens}\ \emph {et~al.}(2006)\citenamefont
  {Claessens}, \citenamefont {Tharmann}, \citenamefont {Kroy},\ and\
  \citenamefont {Bausch}}]{Claessens:2006}%
  \BibitemOpen
  \bibfield  {author} {\bibinfo {author} {\bibfnamefont {M.}~\bibnamefont
  {Claessens}}, \bibinfo {author} {\bibfnamefont {R.}~\bibnamefont {Tharmann}},
  \bibinfo {author} {\bibfnamefont {K.}~\bibnamefont {Kroy}}, \ and\ \bibinfo
  {author} {\bibfnamefont {A.}~\bibnamefont {Bausch}},\ }\href
  {http://dx.doi.org/10.1038/nphys241} {\bibfield  {journal} {\bibinfo
  {journal} {Nature Physics}\ }\textbf {\bibinfo {volume} {2}},\ \bibinfo
  {pages} {186} (\bibinfo {year} {2006})}\BibitemShut {NoStop}%
\bibitem [{\citenamefont {Chaudhuri}\ \emph {et~al.}(2007)\citenamefont
  {Chaudhuri}, \citenamefont {Parekh},\ and\ \citenamefont
  {Fletcher}}]{Chaudhuri:2007}%
  \BibitemOpen
  \bibfield  {author} {\bibinfo {author} {\bibfnamefont {O.}~\bibnamefont
  {Chaudhuri}}, \bibinfo {author} {\bibfnamefont {S.~H.}\ \bibnamefont
  {Parekh}}, \ and\ \bibinfo {author} {\bibfnamefont {D.~A.}\ \bibnamefont
  {Fletcher}},\ }\href {\doibase 10.1038/nature05459} {\bibfield  {journal}
  {\bibinfo  {journal} {Nature}\ }\textbf {\bibinfo {volume} {445}},\ \bibinfo
  {pages} {295} (\bibinfo {year} {2007})}\BibitemShut {NoStop}%
\bibitem [{\citenamefont {Huisman}\ \emph {et~al.}(2008)\citenamefont
  {Huisman}, \citenamefont {Storm},\ and\ \citenamefont
  {Barkema}}]{Huisman:2008}%
  \BibitemOpen
  \bibfield  {author} {\bibinfo {author} {\bibfnamefont {E.~M.}\ \bibnamefont
  {Huisman}}, \bibinfo {author} {\bibfnamefont {C.}~\bibnamefont {Storm}}, \
  and\ \bibinfo {author} {\bibfnamefont {G.~T.}\ \bibnamefont {Barkema}},\
  }\href {\doibase 10.1103/PhysRevE.78.051801} {\bibfield  {journal} {\bibinfo
  {journal} {Phys. Rev. E}\ }\textbf {\bibinfo {volume} {78}},\ \bibinfo
  {pages} {051801} (\bibinfo {year} {2008})}\BibitemShut {NoStop}%
\bibitem [{\citenamefont {Carrillo}\ \emph {et~al.}(2013)\citenamefont
  {Carrillo}, \citenamefont {MacKintosh},\ and\ \citenamefont
  {Dobrynin}}]{Carillo:2013}%
  \BibitemOpen
  \bibfield  {author} {\bibinfo {author} {\bibfnamefont {J.-M.~Y.}\
  \bibnamefont {Carrillo}}, \bibinfo {author} {\bibfnamefont {F.~C.}\
  \bibnamefont {MacKintosh}}, \ and\ \bibinfo {author} {\bibfnamefont {A.~V.}\
  \bibnamefont {Dobrynin}},\ }\href {\doibase 10.1021/ma400478f} {\bibfield
  {journal} {\bibinfo  {journal} {Macromolecules}\ }\textbf {\bibinfo {volume}
  {46}},\ \bibinfo {pages} {3679} (\bibinfo {year} {2013})}\BibitemShut
  {NoStop}%
\bibitem [{\citenamefont {Razbin}\ \emph {et~al.}(2015)\citenamefont {Razbin},
  \citenamefont {Falcke}, \citenamefont {Benetatos},\ and\ \citenamefont
  {Zippelius}}]{Razbin:2015}%
  \BibitemOpen
  \bibfield  {author} {\bibinfo {author} {\bibfnamefont {M.}~\bibnamefont
  {Razbin}}, \bibinfo {author} {\bibfnamefont {M.}~\bibnamefont {Falcke}},
  \bibinfo {author} {\bibfnamefont {P.}~\bibnamefont {Benetatos}}, \ and\
  \bibinfo {author} {\bibfnamefont {A.}~\bibnamefont {Zippelius}},\ }\href
  {http://stacks.iop.org/1478-3975/12/i=4/a=046007} {\bibfield  {journal}
  {\bibinfo  {journal} {Physical Biology}\ }\textbf {\bibinfo {volume} {12}},\
  \bibinfo {pages} {046007} (\bibinfo {year} {2015})}\BibitemShut {NoStop}%
\bibitem [{\citenamefont {Amuasi}\ \emph {et~al.}(2015)\citenamefont {Amuasi},
  \citenamefont {Heussinger}, \citenamefont {Vink},\ and\ \citenamefont
  {Zippelius}}]{Amuasi:2015}%
  \BibitemOpen
  \bibfield  {author} {\bibinfo {author} {\bibfnamefont {H.}~\bibnamefont
  {Amuasi}}, \bibinfo {author} {\bibfnamefont {C.}~\bibnamefont {Heussinger}},
  \bibinfo {author} {\bibfnamefont {R.}~\bibnamefont {Vink}}, \ and\ \bibinfo
  {author} {\bibfnamefont {A.}~\bibnamefont {Zippelius}},\ }\href
  {http://stacks.iop.org/1367-2630/17/i=8/a=083035} {\bibfield  {journal}
  {\bibinfo  {journal} {New Journal of Physics}\ }\textbf {\bibinfo {volume}
  {17}},\ \bibinfo {pages} {083035} (\bibinfo {year} {2015})}\BibitemShut
  {NoStop}%
\bibitem [{\citenamefont {Plagge}\ \emph {et~al.}(2016)\citenamefont {Plagge},
  \citenamefont {Fischer},\ and\ \citenamefont {Heussinger}}]{Plagge:2016}%
  \BibitemOpen
  \bibfield  {author} {\bibinfo {author} {\bibfnamefont {J.}~\bibnamefont
  {Plagge}}, \bibinfo {author} {\bibfnamefont {A.}~\bibnamefont {Fischer}}, \
  and\ \bibinfo {author} {\bibfnamefont {C.}~\bibnamefont {Heussinger}},\
  }\href {\doibase 10.1103/PhysRevE.93.062502} {\bibfield  {journal} {\bibinfo
  {journal} {Phys. Rev. E}\ }\textbf {\bibinfo {volume} {93}},\ \bibinfo
  {pages} {062502} (\bibinfo {year} {2016})}\BibitemShut {NoStop}%
\bibitem [{\citenamefont {Ratner}\ \emph {et~al.}(2004)\citenamefont {Ratner},
  \citenamefont {Hoffman}, \citenamefont {Schoen},\ and\ \citenamefont
  {Lemons}}]{Ratner:2004}%
  \BibitemOpen
  \bibfield  {author} {\bibinfo {author} {\bibfnamefont {B.}~\bibnamefont
  {Ratner}}, \bibinfo {author} {\bibfnamefont {A.}~\bibnamefont {Hoffman}},
  \bibinfo {author} {\bibfnamefont {F.}~\bibnamefont {Schoen}}, \ and\ \bibinfo
  {author} {\bibfnamefont {J.}~\bibnamefont {Lemons}},\ }\href
  {https://books.google.at/books?id=9PMU1iYGe34C} {\emph {\bibinfo {title}
  {Biomaterials Science: An Introduction to Materials in Medicine}}}\ (\bibinfo
   {publisher} {Elsevier Academic Press, San Diego, London},\ \bibinfo {year}
  {2004})\BibitemShut {NoStop}%
\bibitem [{\citenamefont {Ratner}\ and\ \citenamefont
  {Bryant}(2004)}]{Ratner:2004:review}%
  \BibitemOpen
  \bibfield  {author} {\bibinfo {author} {\bibfnamefont {B.~D.}\ \bibnamefont
  {Ratner}}\ and\ \bibinfo {author} {\bibfnamefont {S.~J.}\ \bibnamefont
  {Bryant}},\ }\href {\doibase 10.1146/annurev.bioeng.6.040803.140027}
  {\bibfield  {journal} {\bibinfo  {journal} {Annual Review of Biomedical
  Engineering}\ }\textbf {\bibinfo {volume} {6}},\ \bibinfo {pages} {41}
  (\bibinfo {year} {2004})}\BibitemShut {NoStop}%
\bibitem [{\citenamefont {Hugel}\ and\ \citenamefont
  {Seitz}(2001)}]{Hugel:2001}%
  \BibitemOpen
  \bibfield  {author} {\bibinfo {author} {\bibfnamefont {T.}~\bibnamefont
  {Hugel}}\ and\ \bibinfo {author} {\bibfnamefont {M.}~\bibnamefont {Seitz}},\
  }\href {\doibase
  10.1002/1521-3927(20010901)22:13<989::AID-MARC989>3.0.CO;2-D} {\bibfield
  {journal} {\bibinfo  {journal} {Macromolecular Rapid Communications}\
  }\textbf {\bibinfo {volume} {22}},\ \bibinfo {pages} {989} (\bibinfo {year}
  {2001})}\BibitemShut {NoStop}%
\bibitem [{\citenamefont {Ott}\ \emph {et~al.}(1993)\citenamefont {Ott},
  \citenamefont {Magnasco}, \citenamefont {Simon},\ and\ \citenamefont
  {Libchaber}}]{Magnasco:1993}%
  \BibitemOpen
  \bibfield  {author} {\bibinfo {author} {\bibfnamefont {A.}~\bibnamefont
  {Ott}}, \bibinfo {author} {\bibfnamefont {M.}~\bibnamefont {Magnasco}},
  \bibinfo {author} {\bibfnamefont {A.}~\bibnamefont {Simon}}, \ and\ \bibinfo
  {author} {\bibfnamefont {A.}~\bibnamefont {Libchaber}},\ }\href {\doibase
  10.1103/PhysRevE.48.R1642} {\bibfield  {journal} {\bibinfo  {journal} {Phys.
  Rev. E}\ }\textbf {\bibinfo {volume} {48}},\ \bibinfo {pages} {R1642}
  (\bibinfo {year} {1993})}\BibitemShut {NoStop}%
\bibitem [{\citenamefont {Broedersz}\ and\ \citenamefont
  {MacKintosh}(2014)}]{MacKintosh:2014}%
  \BibitemOpen
  \bibfield  {author} {\bibinfo {author} {\bibfnamefont {C.~P.}\ \bibnamefont
  {Broedersz}}\ and\ \bibinfo {author} {\bibfnamefont {F.~C.}\ \bibnamefont
  {MacKintosh}},\ }\href {\doibase 10.1103/RevModPhys.86.995} {\bibfield
  {journal} {\bibinfo  {journal} {Rev. Mod. Phys.}\ }\textbf {\bibinfo {volume}
  {86}},\ \bibinfo {pages} {995} (\bibinfo {year} {2014})}\BibitemShut
  {NoStop}%
\bibitem [{\citenamefont {Razbin}\ \emph {et~al.}(2016)\citenamefont {Razbin},
  \citenamefont {Benetatos},\ and\ \citenamefont {Zippelius}}]{Razbin:2016}%
  \BibitemOpen
  \bibfield  {author} {\bibinfo {author} {\bibfnamefont {M.}~\bibnamefont
  {Razbin}}, \bibinfo {author} {\bibfnamefont {P.}~\bibnamefont {Benetatos}}, \
  and\ \bibinfo {author} {\bibfnamefont {A.}~\bibnamefont {Zippelius}},\ }\href
  {\doibase 10.1103/PhysRevE.93.052408} {\bibfield  {journal} {\bibinfo
  {journal} {Phys. Rev. E}\ }\textbf {\bibinfo {volume} {93}},\ \bibinfo
  {pages} {052408} (\bibinfo {year} {2016})}\BibitemShut {NoStop}%
\bibitem [{\citenamefont {Ashkin}(1997)}]{Ashkin:1997}%
  \BibitemOpen
  \bibfield  {author} {\bibinfo {author} {\bibfnamefont {A.}~\bibnamefont
  {Ashkin}},\ }\href {http://www.pnas.org/content/94/10/4853.abstract}
  {\bibfield  {journal} {\bibinfo  {journal} {Proceedings of the National
  Academy of Sciences}\ }\textbf {\bibinfo {volume} {94}},\ \bibinfo {pages}
  {4853} (\bibinfo {year} {1997})}\BibitemShut {NoStop}%
\bibitem [{\citenamefont {Mehta}\ \emph {et~al.}(1999)\citenamefont {Mehta},
  \citenamefont {Rief}, \citenamefont {Spudich}, \citenamefont {Smith},\ and\
  \citenamefont {Simmons}}]{Mehta:1999}%
  \BibitemOpen
  \bibfield  {author} {\bibinfo {author} {\bibfnamefont {A.~D.}\ \bibnamefont
  {Mehta}}, \bibinfo {author} {\bibfnamefont {M.}~\bibnamefont {Rief}},
  \bibinfo {author} {\bibfnamefont {J.~A.}\ \bibnamefont {Spudich}}, \bibinfo
  {author} {\bibfnamefont {D.~A.}\ \bibnamefont {Smith}}, \ and\ \bibinfo
  {author} {\bibfnamefont {R.~M.}\ \bibnamefont {Simmons}},\ }\href {\doibase
  10.1126/science.283.5408.1689} {\bibfield  {journal} {\bibinfo  {journal}
  {Science}\ }\textbf {\bibinfo {volume} {283}},\ \bibinfo {pages} {1689}
  (\bibinfo {year} {1999})}\BibitemShut {NoStop}%
\bibitem [{\citenamefont {Gosse}\ and\ \citenamefont
  {Croquette}(2002)}]{Gosse:2002}%
  \BibitemOpen
  \bibfield  {author} {\bibinfo {author} {\bibfnamefont {C.}~\bibnamefont
  {Gosse}}\ and\ \bibinfo {author} {\bibfnamefont {V.}~\bibnamefont
  {Croquette}},\ }\href {\doibase 10.1016/S0006-3495(02)75672-5} {\bibfield
  {journal} {\bibinfo  {journal} {Biophysical Journal}\ }\textbf {\bibinfo
  {volume} {82}},\ \bibinfo {pages} {3314 } (\bibinfo {year}
  {2002})}\BibitemShut {NoStop}%
\bibitem [{\citenamefont {Kuzumaki}\ and\ \citenamefont
  {Mitsuda}(2006)}]{Kuzumaki:2006}%
  \BibitemOpen
  \bibfield  {author} {\bibinfo {author} {\bibfnamefont {T.}~\bibnamefont
  {Kuzumaki}}\ and\ \bibinfo {author} {\bibfnamefont {Y.}~\bibnamefont
  {Mitsuda}},\ }\href {http://stacks.iop.org/1347-4065/45/i=1R/a=364}
  {\bibfield  {journal} {\bibinfo  {journal} {Japanese Journal of Applied
  Physics}\ }\textbf {\bibinfo {volume} {45}},\ \bibinfo {pages} {364}
  (\bibinfo {year} {2006})}\BibitemShut {NoStop}%
\bibitem [{\citenamefont {Janshoff}\ \emph {et~al.}(2000)\citenamefont
  {Janshoff}, \citenamefont {Neitzert}, \citenamefont {Oberd\"orfer},\ and\
  \citenamefont {Fuchs}}]{Janshoff:2000}%
  \BibitemOpen
  \bibfield  {author} {\bibinfo {author} {\bibfnamefont {A.}~\bibnamefont
  {Janshoff}}, \bibinfo {author} {\bibfnamefont {M.}~\bibnamefont {Neitzert}},
  \bibinfo {author} {\bibfnamefont {Y.}~\bibnamefont {Oberd\"orfer}}, \ and\
  \bibinfo {author} {\bibfnamefont {H.}~\bibnamefont {Fuchs}},\ }\href
  {\doibase 10.1002/1521-3773(20000915)39:18<3212::AID-ANIE3212>3.0.CO;2-X}
  {\bibfield  {journal} {\bibinfo  {journal} {Angewandte Chemie International
  Edition}\ }\textbf {\bibinfo {volume} {39}},\ \bibinfo {pages} {3212}
  (\bibinfo {year} {2000})}\BibitemShut {NoStop}%
\bibitem [{\citenamefont {Saleh}(2015)}]{Saleh:2015}%
  \BibitemOpen
  \bibfield  {author} {\bibinfo {author} {\bibfnamefont {O.~A.}\ \bibnamefont
  {Saleh}},\ }\href {\doibase 10.1063/1.4921348} {\bibfield  {journal}
  {\bibinfo  {journal} {The Journal of chemical physics}\ }\textbf {\bibinfo
  {volume} {142}},\ \bibinfo {pages} {194902} (\bibinfo {year}
  {2015})}\BibitemShut {NoStop}%
\bibitem [{\citenamefont {Marko}\ and\ \citenamefont
  {Siggia}(1995)}]{Marko:1995}%
  \BibitemOpen
  \bibfield  {author} {\bibinfo {author} {\bibfnamefont {J.~F.}\ \bibnamefont
  {Marko}}\ and\ \bibinfo {author} {\bibfnamefont {E.~D.}\ \bibnamefont
  {Siggia}},\ }\href {\doibase 10.1021/ma00130a008} {\bibfield  {journal}
  {\bibinfo  {journal} {Macromolecules}\ }\textbf {\bibinfo {volume} {28}},\
  \bibinfo {pages} {8759} (\bibinfo {year} {1995})}\BibitemShut {NoStop}%
\bibitem [{\citenamefont {Bouchiat}\ \emph {et~al.}(1999)\citenamefont
  {Bouchiat}, \citenamefont {Wang}, \citenamefont {Allemand}, \citenamefont
  {Strick}, \citenamefont {Block},\ and\ \citenamefont
  {Croquette}}]{Bouchiat:1999}%
  \BibitemOpen
  \bibfield  {author} {\bibinfo {author} {\bibfnamefont {C.}~\bibnamefont
  {Bouchiat}}, \bibinfo {author} {\bibfnamefont {M.}~\bibnamefont {Wang}},
  \bibinfo {author} {\bibfnamefont {J.-F.}\ \bibnamefont {Allemand}}, \bibinfo
  {author} {\bibfnamefont {T.}~\bibnamefont {Strick}}, \bibinfo {author}
  {\bibfnamefont {S.}~\bibnamefont {Block}}, \ and\ \bibinfo {author}
  {\bibfnamefont {V.}~\bibnamefont {Croquette}},\ }\href {\doibase
  10.1016/S0006-3495(99)77207-3} {\bibfield  {journal} {\bibinfo  {journal}
  {Biophysical Journal}\ }\textbf {\bibinfo {volume} {76}},\ \bibinfo {pages}
  {409 } (\bibinfo {year} {1999})}\BibitemShut {NoStop}%
\bibitem [{\citenamefont {Bustamante}\ \emph {et~al.}(2000)\citenamefont
  {Bustamante}, \citenamefont {Smith}, \citenamefont {Liphardt},\ and\
  \citenamefont {Smith}}]{Bustamante:2000}%
  \BibitemOpen
  \bibfield  {author} {\bibinfo {author} {\bibfnamefont {C.}~\bibnamefont
  {Bustamante}}, \bibinfo {author} {\bibfnamefont {S.~B.}\ \bibnamefont
  {Smith}}, \bibinfo {author} {\bibfnamefont {J.}~\bibnamefont {Liphardt}}, \
  and\ \bibinfo {author} {\bibfnamefont {D.}~\bibnamefont {Smith}},\ }\href
  {\doibase 10.1016/S0959-440X(00)00085-3} {\bibfield  {journal} {\bibinfo
  {journal} {Current Opinion in Structural Biology}\ }\textbf {\bibinfo
  {volume} {10}},\ \bibinfo {pages} {279 } (\bibinfo {year}
  {2000})}\BibitemShut {NoStop}%
\bibitem [{\citenamefont {Liu}\ and\ \citenamefont {Pollack}(2002)}]{Liu:2002}%
  \BibitemOpen
  \bibfield  {author} {\bibinfo {author} {\bibfnamefont {X.}~\bibnamefont
  {Liu}}\ and\ \bibinfo {author} {\bibfnamefont {G.~H.}\ \bibnamefont
  {Pollack}},\ }\href {\doibase 10.1016/S0006-3495(02)75280-6} {\bibfield
  {journal} {\bibinfo  {journal} {Biophysics Journal}\ }\textbf {\bibinfo
  {volume} {83}},\ \bibinfo {pages} {2705} (\bibinfo {year}
  {2002})}\BibitemShut {NoStop}%
\bibitem [{\citenamefont {Kellermayer}\ \emph {et~al.}(1997)\citenamefont
  {Kellermayer}, \citenamefont {Smith}, \citenamefont {Granzier},\ and\
  \citenamefont {Bustamante}}]{Kellermayer:1997}%
  \BibitemOpen
  \bibfield  {author} {\bibinfo {author} {\bibfnamefont {M.~S.~Z.}\
  \bibnamefont {Kellermayer}}, \bibinfo {author} {\bibfnamefont {S.~B.}\
  \bibnamefont {Smith}}, \bibinfo {author} {\bibfnamefont {H.~L.}\ \bibnamefont
  {Granzier}}, \ and\ \bibinfo {author} {\bibfnamefont {C.}~\bibnamefont
  {Bustamante}},\ }\href {\doibase 10.1126/science.276.5315.1112} {\bibfield
  {journal} {\bibinfo  {journal} {Science}\ }\textbf {\bibinfo {volume}
  {276}},\ \bibinfo {pages} {1112} (\bibinfo {year} {1997})}\BibitemShut
  {NoStop}%
\bibitem [{\citenamefont {Sun}\ \emph {et~al.}(2002)\citenamefont {Sun},
  \citenamefont {Luo}, \citenamefont {Fertala},\ and\ \citenamefont
  {An}}]{Sun:2002}%
  \BibitemOpen
  \bibfield  {author} {\bibinfo {author} {\bibfnamefont {Y.-L.}\ \bibnamefont
  {Sun}}, \bibinfo {author} {\bibfnamefont {Z.-P.}\ \bibnamefont {Luo}},
  \bibinfo {author} {\bibfnamefont {A.}~\bibnamefont {Fertala}}, \ and\
  \bibinfo {author} {\bibfnamefont {K.-N.}\ \bibnamefont {An}},\ }\href
  {\doibase 10.1016/S0006-291X(02)00685-X} {\bibfield  {journal} {\bibinfo
  {journal} {Biochemical and Biophysical Research Communications}\ }\textbf
  {\bibinfo {volume} {295}},\ \bibinfo {pages} {382 } (\bibinfo {year}
  {2002})}\BibitemShut {NoStop}%
\bibitem [{\citenamefont {Kratky}\ and\ \citenamefont
  {Porod}(1949)}]{Kratky:1949}%
  \BibitemOpen
  \bibfield  {author} {\bibinfo {author} {\bibfnamefont {O.}~\bibnamefont
  {Kratky}}\ and\ \bibinfo {author} {\bibfnamefont {G.}~\bibnamefont {Porod}},\
  }\href {\doibase 10.1002/recl.19490681203} {\bibfield  {journal} {\bibinfo
  {journal} {Recueil des Travaux Chimiques des Pays-Bas}\ }\textbf {\bibinfo
  {volume} {68}},\ \bibinfo {pages} {1106} (\bibinfo {year}
  {1949})}\BibitemShut {NoStop}%
\bibitem [{\citenamefont {Wilhelm}\ and\ \citenamefont
  {Frey}(1996)}]{Wilhelm:1996}%
  \BibitemOpen
  \bibfield  {author} {\bibinfo {author} {\bibfnamefont {J.}~\bibnamefont
  {Wilhelm}}\ and\ \bibinfo {author} {\bibfnamefont {E.}~\bibnamefont {Frey}},\
  }\href {\doibase 10.1103/PhysRevLett.77.2581} {\bibfield  {journal} {\bibinfo
   {journal} {Phys. Rev. Lett.}\ }\textbf {\bibinfo {volume} {77}},\ \bibinfo
  {pages} {2581} (\bibinfo {year} {1996})}\BibitemShut {NoStop}%
\bibitem [{\citenamefont {Hamprecht}\ \emph {et~al.}(2004)\citenamefont
  {Hamprecht}, \citenamefont {Janke},\ and\ \citenamefont
  {Kleinert}}]{Hamprecht:2004}%
  \BibitemOpen
  \bibfield  {author} {\bibinfo {author} {\bibfnamefont {B.}~\bibnamefont
  {Hamprecht}}, \bibinfo {author} {\bibfnamefont {W.}~\bibnamefont {Janke}}, \
  and\ \bibinfo {author} {\bibfnamefont {H.}~\bibnamefont {Kleinert}},\ }\href
  {\doibase 10.1016/j.physleta.2004.06.104} {\bibfield  {journal} {\bibinfo
  {journal} {Physics Letters A}\ }\textbf {\bibinfo {volume} {330}},\ \bibinfo
  {pages} {254 } (\bibinfo {year} {2004})}\BibitemShut {NoStop}%
\bibitem [{\citenamefont {Spakowitz}\ \emph {et~al.}(2004)\citenamefont
  {Spakowitz}, ,\ and\ \citenamefont {Wang}}]{Spakowitz:2004}%
  \BibitemOpen
  \bibfield  {author} {\bibinfo {author} {\bibfnamefont {A.~J.}\ \bibnamefont
  {Spakowitz}}, , \ and\ \bibinfo {author} {\bibfnamefont {Z.-G.}\ \bibnamefont
  {Wang}},\ }\href {\doibase 10.1021/ma049958v} {\bibfield  {journal} {\bibinfo
   {journal} {Macromolecules}\ }\textbf {\bibinfo {volume} {37}},\ \bibinfo
  {pages} {5814} (\bibinfo {year} {2004})}\BibitemShut {NoStop}%
\bibitem [{\citenamefont {Spakowitz}\ and\ \citenamefont
  {Wang}(2005)}]{Spakowitz:2005}%
  \BibitemOpen
  \bibfield  {author} {\bibinfo {author} {\bibfnamefont {A.~J.}\ \bibnamefont
  {Spakowitz}}\ and\ \bibinfo {author} {\bibfnamefont {Z.-G.}\ \bibnamefont
  {Wang}},\ }\href {\doibase 10.1103/PhysRevE.72.041802} {\bibfield  {journal}
  {\bibinfo  {journal} {Phys. Rev. E}\ }\textbf {\bibinfo {volume} {72}},\
  \bibinfo {pages} {041802} (\bibinfo {year} {2005})}\BibitemShut {NoStop}%
\bibitem [{\citenamefont {Mehraeen}\ \emph {et~al.}(2008)\citenamefont
  {Mehraeen}, \citenamefont {Sudhanshu}, \citenamefont {Koslover},\ and\
  \citenamefont {Spakowitz}}]{Mehraeen:2008}%
  \BibitemOpen
  \bibfield  {author} {\bibinfo {author} {\bibfnamefont {S.}~\bibnamefont
  {Mehraeen}}, \bibinfo {author} {\bibfnamefont {B.}~\bibnamefont {Sudhanshu}},
  \bibinfo {author} {\bibfnamefont {E.~F.}\ \bibnamefont {Koslover}}, \ and\
  \bibinfo {author} {\bibfnamefont {A.~J.}\ \bibnamefont {Spakowitz}},\ }\href
  {\doibase 10.1103/PhysRevE.77.061803} {\bibfield  {journal} {\bibinfo
  {journal} {Phys. Rev. E}\ }\textbf {\bibinfo {volume} {77}},\ \bibinfo
  {pages} {061803} (\bibinfo {year} {2008})}\BibitemShut {NoStop}%
\bibitem [{\citenamefont {Taute}\ \emph {et~al.}(2008)\citenamefont {Taute},
  \citenamefont {Pampaloni}, \citenamefont {Frey},\ and\ \citenamefont
  {Florin}}]{Taute:2008}%
  \BibitemOpen
  \bibfield  {author} {\bibinfo {author} {\bibfnamefont {K.~M.}\ \bibnamefont
  {Taute}}, \bibinfo {author} {\bibfnamefont {F.}~\bibnamefont {Pampaloni}},
  \bibinfo {author} {\bibfnamefont {E.}~\bibnamefont {Frey}}, \ and\ \bibinfo
  {author} {\bibfnamefont {E.-L.}\ \bibnamefont {Florin}},\ }\href {\doibase
  10.1103/PhysRevLett.100.028102} {\bibfield  {journal} {\bibinfo  {journal}
  {Phys. Rev. Lett.}\ }\textbf {\bibinfo {volume} {100}},\ \bibinfo {pages}
  {028102} (\bibinfo {year} {2008})}\BibitemShut {NoStop}%
\bibitem [{\citenamefont {Pampaloni}\ \emph {et~al.}(2006)\citenamefont
  {Pampaloni}, \citenamefont {Lattanzi}, \citenamefont {Jon{\'a}{\v{s}}},
  \citenamefont {Surrey}, \citenamefont {Frey},\ and\ \citenamefont
  {Florin}}]{Pampaloni:2006}%
  \BibitemOpen
  \bibfield  {author} {\bibinfo {author} {\bibfnamefont {F.}~\bibnamefont
  {Pampaloni}}, \bibinfo {author} {\bibfnamefont {G.}~\bibnamefont {Lattanzi}},
  \bibinfo {author} {\bibfnamefont {A.}~\bibnamefont {Jon{\'a}{\v{s}}}},
  \bibinfo {author} {\bibfnamefont {T.}~\bibnamefont {Surrey}}, \bibinfo
  {author} {\bibfnamefont {E.}~\bibnamefont {Frey}}, \ and\ \bibinfo {author}
  {\bibfnamefont {E.-L.}\ \bibnamefont {Florin}},\ }\href {\doibase
  10.1073/pnas.0603931103} {\bibfield  {journal} {\bibinfo  {journal}
  {Proceedings of the National Academy of Sciences}\ }\textbf {\bibinfo
  {volume} {103}},\ \bibinfo {pages} {10248} (\bibinfo {year}
  {2006})}\BibitemShut {NoStop}%
\bibitem [{\citenamefont {Prasad}\ \emph {et~al.}(2005)\citenamefont {Prasad},
  \citenamefont {Hori},\ and\ \citenamefont {Kondev}}]{Prasad:2005}%
  \BibitemOpen
  \bibfield  {author} {\bibinfo {author} {\bibfnamefont {A.}~\bibnamefont
  {Prasad}}, \bibinfo {author} {\bibfnamefont {Y.}~\bibnamefont {Hori}}, \ and\
  \bibinfo {author} {\bibfnamefont {J.}~\bibnamefont {Kondev}},\ }\href
  {\doibase 10.1103/PhysRevE.72.041918} {\bibfield  {journal} {\bibinfo
  {journal} {Phys. Rev. E}\ }\textbf {\bibinfo {volume} {72}},\ \bibinfo
  {pages} {041918} (\bibinfo {year} {2005})}\BibitemShut {NoStop}%
\bibitem [{\citenamefont {Baczynski}\ \emph {et~al.}(2007)\citenamefont
  {Baczynski}, \citenamefont {Lipowsky},\ and\ \citenamefont
  {Kierfeld}}]{Baczynski:2007}%
  \BibitemOpen
  \bibfield  {author} {\bibinfo {author} {\bibfnamefont {K.}~\bibnamefont
  {Baczynski}}, \bibinfo {author} {\bibfnamefont {R.}~\bibnamefont {Lipowsky}},
  \ and\ \bibinfo {author} {\bibfnamefont {J.}~\bibnamefont {Kierfeld}},\
  }\href {\doibase 10.1103/PhysRevE.76.061914} {\bibfield  {journal} {\bibinfo
  {journal} {Phys. Rev. E}\ }\textbf {\bibinfo {volume} {76}},\ \bibinfo
  {pages} {061914} (\bibinfo {year} {2007})}\BibitemShut {NoStop}%
\bibitem [{\citenamefont {Emanuel}\ \emph {et~al.}(2007)\citenamefont
  {Emanuel}, \citenamefont {Mohrbach}, \citenamefont {Sayar}, \citenamefont
  {Schiessel},\ and\ \citenamefont {Kuli\ifmmode~\acute{c}\else
  \'{c}\fi{}}}]{Emanuel:2007}%
  \BibitemOpen
  \bibfield  {author} {\bibinfo {author} {\bibfnamefont {M.}~\bibnamefont
  {Emanuel}}, \bibinfo {author} {\bibfnamefont {H.}~\bibnamefont {Mohrbach}},
  \bibinfo {author} {\bibfnamefont {M.}~\bibnamefont {Sayar}}, \bibinfo
  {author} {\bibfnamefont {H.}~\bibnamefont {Schiessel}}, \ and\ \bibinfo
  {author} {\bibfnamefont {I.~M.}\ \bibnamefont {Kuli\ifmmode~\acute{c}\else
  \'{c}\fi{}}},\ }\href {\doibase 10.1103/PhysRevE.76.061907} {\bibfield
  {journal} {\bibinfo  {journal} {Phys. Rev. E}\ }\textbf {\bibinfo {volume}
  {76}},\ \bibinfo {pages} {061907} (\bibinfo {year} {2007})}\BibitemShut
  {NoStop}%
\bibitem [{\citenamefont {Lee}\ \emph {et~al.}(2007)\citenamefont {Lee},
  \citenamefont {Johner},\ and\ \citenamefont {Hong}}]{Lee:2007}%
  \BibitemOpen
  \bibfield  {author} {\bibinfo {author} {\bibfnamefont {N.~K.}\ \bibnamefont
  {Lee}}, \bibinfo {author} {\bibfnamefont {A.}~\bibnamefont {Johner}}, \ and\
  \bibinfo {author} {\bibfnamefont {S.~C.}\ \bibnamefont {Hong}},\ }\href
  {\doibase 10.1140/epje/i2007-10230-4} {\bibfield  {journal} {\bibinfo
  {journal} {The European Physical Journal E}\ }\textbf {\bibinfo {volume}
  {24}},\ \bibinfo {pages} {229} (\bibinfo {year} {2007})}\BibitemShut
  {NoStop}%
\bibitem [{\citenamefont {Bedi}\ and\ \citenamefont {Mao}(2015)}]{Bedi:2015}%
  \BibitemOpen
  \bibfield  {author} {\bibinfo {author} {\bibfnamefont {D.~S.}\ \bibnamefont
  {Bedi}}\ and\ \bibinfo {author} {\bibfnamefont {X.}~\bibnamefont {Mao}},\
  }\href {\doibase 10.1103/PhysRevE.92.062141} {\bibfield  {journal} {\bibinfo
  {journal} {Phys. Rev. E}\ }\textbf {\bibinfo {volume} {92}},\ \bibinfo
  {pages} {062141} (\bibinfo {year} {2015})}\BibitemShut {NoStop}%
\bibitem [{\citenamefont {Ghosh}\ \emph {et~al.}(2007)\citenamefont {Ghosh},
  \citenamefont {Samuel},\ and\ \citenamefont {Sinha}}]{Ghosh:2007}%
  \BibitemOpen
  \bibfield  {author} {\bibinfo {author} {\bibfnamefont {A.}~\bibnamefont
  {Ghosh}}, \bibinfo {author} {\bibfnamefont {J.}~\bibnamefont {Samuel}}, \
  and\ \bibinfo {author} {\bibfnamefont {S.}~\bibnamefont {Sinha}},\ }\href
  {\doibase 10.1103/PhysRevE.76.061801} {\bibfield  {journal} {\bibinfo
  {journal} {Phys. Rev. E}\ }\textbf {\bibinfo {volume} {76}},\ \bibinfo
  {pages} {061801} (\bibinfo {year} {2007})}\BibitemShut {NoStop}%
\bibitem [{\citenamefont {Doi}\ and\ \citenamefont {Edwards}(1986)}]{Doi:1986}%
  \BibitemOpen
  \bibfield  {author} {\bibinfo {author} {\bibfnamefont {M.}~\bibnamefont
  {Doi}}\ and\ \bibinfo {author} {\bibfnamefont {S.~F.}\ \bibnamefont
  {Edwards}},\ }\href {https://books.google.co.in/books?id=JSFsAAAACAAJ} {\emph
  {\bibinfo {title} {The Theory of Polymer Dynamics}}}\ (\bibinfo  {publisher}
  {Oxford Science Publications},\ \bibinfo {year} {1986})\BibitemShut {NoStop}%
\bibitem [{\citenamefont {Landau}\ and\ \citenamefont
  {Lifshitz}(1986)}]{Landau:1986}%
  \BibitemOpen
  \bibfield  {author} {\bibinfo {author} {\bibfnamefont {L.~D.}\ \bibnamefont
  {Landau}}\ and\ \bibinfo {author} {\bibfnamefont {E.}~\bibnamefont
  {Lifshitz}},\ }\href {https://books.google.at/books?id=tpY-VkwCkAIC}
  {\bibfield  {journal} {\bibinfo  {journal} {Course of Theoretical Physics}\
  }\textbf {\bibinfo {volume} {7}} (\bibinfo {year} {1986})}\BibitemShut
  {NoStop}%
\bibitem [{\citenamefont {Kleinert}(2009)}]{Kleinert:2009}%
  \BibitemOpen
  \bibfield  {author} {\bibinfo {author} {\bibfnamefont {H.}~\bibnamefont
  {Kleinert}},\ }\href {https://books.google.at/books?id=VJ1qNz5xYzkC} {\emph
  {\bibinfo {title} {Path Integrals in Quantum Mechanics, Statistics, Polymer
  Physics, and Financial Markets}}},\ EBL-Schweitzer\ (\bibinfo  {publisher}
  {World Scientific, Singapore},\ \bibinfo {year} {2009})\BibitemShut {NoStop}%
\bibitem [{\citenamefont {Aldrovandi}\ and\ \citenamefont
  {Ferreira}(1980)}]{Aldrovandi:1980}%
  \BibitemOpen
  \bibfield  {author} {\bibinfo {author} {\bibfnamefont {R.}~\bibnamefont
  {Aldrovandi}}\ and\ \bibinfo {author} {\bibfnamefont {P.~L.}\ \bibnamefont
  {Ferreira}},\ }\href {\doibase 10.1119/1.12332} {\bibfield  {journal}
  {\bibinfo  {journal} {American Journal of Physics}\ }\textbf {\bibinfo
  {volume} {48}},\ \bibinfo {pages} {660} (\bibinfo {year} {1980})}\BibitemShut
  {NoStop}%
\bibitem [{{\relax DLMF}()}]{NIST:online}%
  \BibitemOpen
  {\relax DLMF},\ \href {http://dlmf.nist.gov/} {\enquote {\bibinfo {title}
  {{NIST Digital Library of Mathematical Functions}},}\ }\bibinfo
  {howpublished} {http://dlmf.nist.gov/, Release 1.0.13 of 2016-09-16},\
  \bibinfo {note} {online companion to \cite{NIST:print}}\BibitemShut {NoStop}%
\bibitem [{\citenamefont {Olver}\ \emph {et~al.}(2010)\citenamefont {Olver},
  \citenamefont {Lozier}, \citenamefont {Boisvert},\ and\ \citenamefont
  {Clark}}]{NIST:print}%
  \BibitemOpen
  \bibinfo {editor} {\bibfnamefont {F.~W.~J.}\ \bibnamefont {Olver}}, \bibinfo
  {editor} {\bibfnamefont {D.~W.}\ \bibnamefont {Lozier}}, \bibinfo {editor}
  {\bibfnamefont {R.~F.}\ \bibnamefont {Boisvert}}, \ and\ \bibinfo {editor}
  {\bibfnamefont {C.~W.}\ \bibnamefont {Clark}},\ eds.,\ \href@noop {} {\emph
  {\bibinfo {title} {{NIST Handbook of Mathematical Functions}}}}\ (\bibinfo
  {publisher} {Cambridge University Press},\ \bibinfo {address} {New York,
  NY},\ \bibinfo {year} {2010})\ \bibinfo {note} {print companion to
  \cite{NIST:online}}\BibitemShut {NoStop}%
\bibitem [{\citenamefont {Wolfram~Research}(2016)}]{Mathematica}%
  \BibitemOpen
  \bibfield  {author} {\bibinfo {author} {\bibfnamefont {I.}~\bibnamefont
  {Wolfram~Research}},\ }\href {https://www.wolfram.com/mathematica/} {\enquote
  {\bibinfo {title} {Mathematica 10.4},}\ } (\bibinfo {year}
  {2016})\BibitemShut {NoStop}%
\bibitem [{\citenamefont {Smith}\ \emph {et~al.}(1992)\citenamefont {Smith},
  \citenamefont {Finzi},\ and\ \citenamefont {Bustamante}}]{Smith:1992}%
  \BibitemOpen
  \bibfield  {author} {\bibinfo {author} {\bibfnamefont {S.~B.}\ \bibnamefont
  {Smith}}, \bibinfo {author} {\bibfnamefont {L.}~\bibnamefont {Finzi}}, \ and\
  \bibinfo {author} {\bibfnamefont {C.}~\bibnamefont {Bustamante}},\ }\href
  {http://science.sciencemag.org/content/258/5085/1122} {\bibfield  {journal}
  {\bibinfo  {journal} {Science}\ }\textbf {\bibinfo {volume} {258}},\ \bibinfo
  {pages} {1122} (\bibinfo {year} {1992})}\BibitemShut {NoStop}%
\bibitem [{\citenamefont {Blundell}\ and\ \citenamefont
  {Terentjev}(2009{\natexlab{a}})}]{Blundell:2009}%
  \BibitemOpen
  \bibfield  {author} {\bibinfo {author} {\bibfnamefont {J.~R.}\ \bibnamefont
  {Blundell}}\ and\ \bibinfo {author} {\bibfnamefont {E.~M.}\ \bibnamefont
  {Terentjev}},\ }\href {\doibase 10.1039/B903583D} {\bibfield  {journal}
  {\bibinfo  {journal} {Soft Matter}\ }\textbf {\bibinfo {volume} {5}},\
  \bibinfo {pages} {4015} (\bibinfo {year} {2009}{\natexlab{a}})}\BibitemShut
  {NoStop}%
\bibitem [{\citenamefont {Meng}\ and\ \citenamefont
  {Terentjev}(2017)}]{Meng:2017}%
  \BibitemOpen
  \bibfield  {author} {\bibinfo {author} {\bibfnamefont {F.}~\bibnamefont
  {Meng}}\ and\ \bibinfo {author} {\bibfnamefont {E.~M.}\ \bibnamefont
  {Terentjev}},\ }\href {\doibase 10.3390/polym9020052} {\bibfield  {journal}
  {\bibinfo  {journal} {Polymers}\ }\textbf {\bibinfo {volume} {9}},\ \bibinfo
  {pages} {52} (\bibinfo {year} {2017})}\BibitemShut {NoStop}%
\bibitem [{\citenamefont {Ha}\ and\ \citenamefont
  {Thirumalai}(1997)}]{Ha:1997}%
  \BibitemOpen
  \bibfield  {author} {\bibinfo {author} {\bibfnamefont {B.-Y.}\ \bibnamefont
  {Ha}}\ and\ \bibinfo {author} {\bibfnamefont {D.}~\bibnamefont
  {Thirumalai}},\ }\href {\doibase 10.1063/1.473128} {\bibfield  {journal}
  {\bibinfo  {journal} {The Journal of chemical physics}\ }\textbf {\bibinfo
  {volume} {106}},\ \bibinfo {pages} {4243} (\bibinfo {year}
  {1997})}\BibitemShut {NoStop}%
\bibitem [{\citenamefont {Palmer}\ and\ \citenamefont
  {Boyce}(2008)}]{Palmer:2008}%
  \BibitemOpen
  \bibfield  {author} {\bibinfo {author} {\bibfnamefont {J.~S.}\ \bibnamefont
  {Palmer}}\ and\ \bibinfo {author} {\bibfnamefont {M.~C.}\ \bibnamefont
  {Boyce}},\ }\href {\doibase 10.1016/j.actbio.2007.12.007} {\bibfield
  {journal} {\bibinfo  {journal} {Acta Biomaterialia}\ }\textbf {\bibinfo
  {volume} {4}},\ \bibinfo {pages} {597} (\bibinfo {year} {2008})}\BibitemShut
  {NoStop}%
\bibitem [{\citenamefont {Blundell}\ and\ \citenamefont
  {Terentjev}(2009{\natexlab{b}})}]{Terentjev:2009}%
  \BibitemOpen
  \bibfield  {author} {\bibinfo {author} {\bibfnamefont {J.}~\bibnamefont
  {Blundell}}\ and\ \bibinfo {author} {\bibfnamefont {E.}~\bibnamefont
  {Terentjev}},\ }\href {\doibase 10.1021/ma9004633} {\bibfield  {journal}
  {\bibinfo  {journal} {Macromolecules}\ }\textbf {\bibinfo {volume} {42}},\
  \bibinfo {pages} {5388} (\bibinfo {year} {2009}{\natexlab{b}})}\BibitemShut
  {NoStop}%
\bibitem [{sup()}]{supplement}%
  \BibitemOpen
  \href@noop {} {\ }\BibitemShut {NoStop}%
\bibitem [{\citenamefont {Kurzthaler}\ \emph {et~al.}(2016)\citenamefont
  {Kurzthaler}, \citenamefont {Leitmann},\ and\ \citenamefont
  {Franosch}}]{Kurzthaler:2016}%
  \BibitemOpen
  \bibfield  {author} {\bibinfo {author} {\bibfnamefont {C.}~\bibnamefont
  {Kurzthaler}}, \bibinfo {author} {\bibfnamefont {S.}~\bibnamefont
  {Leitmann}}, \ and\ \bibinfo {author} {\bibfnamefont {T.}~\bibnamefont
  {Franosch}},\ }\href {\doibase 10.1038/srep36702} {\bibfield  {journal}
  {\bibinfo  {journal} {Scientific Reports}\ }\textbf {\bibinfo {volume} {6}},\
  \bibinfo {pages} {36702} (\bibinfo {year} {2016})}\BibitemShut {NoStop}%
\bibitem [{\citenamefont {Leitmann}\ \emph {et~al.}(2016)\citenamefont
  {Leitmann}, \citenamefont {H\"ofling},\ and\ \citenamefont
  {Franosch}}]{Leitmann:2016}%
  \BibitemOpen
  \bibfield  {author} {\bibinfo {author} {\bibfnamefont {S.}~\bibnamefont
  {Leitmann}}, \bibinfo {author} {\bibfnamefont {F.}~\bibnamefont {H\"ofling}},
  \ and\ \bibinfo {author} {\bibfnamefont {T.}~\bibnamefont {Franosch}},\
  }\href {\doibase 10.1103/PhysRevLett.117.097801} {\bibfield  {journal}
  {\bibinfo  {journal} {Phys. Rev. Lett.}\ }\textbf {\bibinfo {volume} {117}},\
  \bibinfo {pages} {097801} (\bibinfo {year} {2016})}\BibitemShut {NoStop}%
\bibitem [{\citenamefont {Sch\"obl}\ \emph {et~al.}(2014)\citenamefont
  {Sch\"obl}, \citenamefont {Sturm}, \citenamefont {Janke},\ and\ \citenamefont
  {Kroy}}]{Schobl:2014}%
  \BibitemOpen
  \bibfield  {author} {\bibinfo {author} {\bibfnamefont {S.}~\bibnamefont
  {Sch\"obl}}, \bibinfo {author} {\bibfnamefont {S.}~\bibnamefont {Sturm}},
  \bibinfo {author} {\bibfnamefont {W.}~\bibnamefont {Janke}}, \ and\ \bibinfo
  {author} {\bibfnamefont {K.}~\bibnamefont {Kroy}},\ }\href {\doibase
  10.1103/PhysRevLett.113.238302} {\bibfield  {journal} {\bibinfo  {journal}
  {Phys. Rev. Lett.}\ }\textbf {\bibinfo {volume} {113}},\ \bibinfo {pages}
  {238302} (\bibinfo {year} {2014})}\BibitemShut {NoStop}%
\bibitem [{\citenamefont {Keshavarz}\ \emph {et~al.}(2016)\citenamefont
  {Keshavarz}, \citenamefont {Engelkamp}, \citenamefont {Xu}, \citenamefont
  {Braeken}, \citenamefont {Otten}, \citenamefont {Uji-i}, \citenamefont
  {Schwartz}, \citenamefont {Koepf}, \citenamefont {Vananroye}, \citenamefont
  {Vermant}, \citenamefont {Nolte}, \citenamefont {De~Schryver}, \citenamefont
  {Maan}, \citenamefont {Hofkens}, \citenamefont {Christianen},\ and\
  \citenamefont {Rowan}}]{Keshavarz:2016}%
  \BibitemOpen
  \bibfield  {author} {\bibinfo {author} {\bibfnamefont {M.}~\bibnamefont
  {Keshavarz}}, \bibinfo {author} {\bibfnamefont {H.}~\bibnamefont
  {Engelkamp}}, \bibinfo {author} {\bibfnamefont {J.}~\bibnamefont {Xu}},
  \bibinfo {author} {\bibfnamefont {E.}~\bibnamefont {Braeken}}, \bibinfo
  {author} {\bibfnamefont {M.~B.~J.}\ \bibnamefont {Otten}}, \bibinfo {author}
  {\bibfnamefont {H.}~\bibnamefont {Uji-i}}, \bibinfo {author} {\bibfnamefont
  {E.}~\bibnamefont {Schwartz}}, \bibinfo {author} {\bibfnamefont
  {M.}~\bibnamefont {Koepf}}, \bibinfo {author} {\bibfnamefont
  {A.}~\bibnamefont {Vananroye}}, \bibinfo {author} {\bibfnamefont
  {J.}~\bibnamefont {Vermant}}, \bibinfo {author} {\bibfnamefont {R.~J.~M.}\
  \bibnamefont {Nolte}}, \bibinfo {author} {\bibfnamefont {F.}~\bibnamefont
  {De~Schryver}}, \bibinfo {author} {\bibfnamefont {J.~C.}\ \bibnamefont
  {Maan}}, \bibinfo {author} {\bibfnamefont {J.}~\bibnamefont {Hofkens}},
  \bibinfo {author} {\bibfnamefont {P.~C.~M.}\ \bibnamefont {Christianen}}, \
  and\ \bibinfo {author} {\bibfnamefont {A.~E.}\ \bibnamefont {Rowan}},\ }\href
  {\doibase 10.1021/acsnano.5b06931} {\bibfield  {journal} {\bibinfo  {journal}
  {ACS Nano}\ }\textbf {\bibinfo {volume} {10}},\ \bibinfo {pages} {1434}
  (\bibinfo {year} {2016})}\BibitemShut {NoStop}%
\bibitem [{\citenamefont {Hallatschek}\ \emph
  {et~al.}(2007{\natexlab{a}})\citenamefont {Hallatschek}, \citenamefont
  {Frey},\ and\ \citenamefont {Kroy}}]{Hallatschek:2007}%
  \BibitemOpen
  \bibfield  {author} {\bibinfo {author} {\bibfnamefont {O.}~\bibnamefont
  {Hallatschek}}, \bibinfo {author} {\bibfnamefont {E.}~\bibnamefont {Frey}}, \
  and\ \bibinfo {author} {\bibfnamefont {K.}~\bibnamefont {Kroy}},\ }\href
  {\doibase 10.1103/PhysRevE.75.031905} {\bibfield  {journal} {\bibinfo
  {journal} {Phys. Rev. E}\ }\textbf {\bibinfo {volume} {75}},\ \bibinfo
  {pages} {031905} (\bibinfo {year} {2007}{\natexlab{a}})}\BibitemShut
  {NoStop}%
\bibitem [{\citenamefont {Hallatschek}\ \emph
  {et~al.}(2007{\natexlab{b}})\citenamefont {Hallatschek}, \citenamefont
  {Frey},\ and\ \citenamefont {Kroy}}]{Hallatschek:2007:2}%
  \BibitemOpen
  \bibfield  {author} {\bibinfo {author} {\bibfnamefont {O.}~\bibnamefont
  {Hallatschek}}, \bibinfo {author} {\bibfnamefont {E.}~\bibnamefont {Frey}}, \
  and\ \bibinfo {author} {\bibfnamefont {K.}~\bibnamefont {Kroy}},\ }\href
  {\doibase 10.1103/PhysRevE.75.031906} {\bibfield  {journal} {\bibinfo
  {journal} {Phys. Rev. E}\ }\textbf {\bibinfo {volume} {75}},\ \bibinfo
  {pages} {031906} (\bibinfo {year} {2007}{\natexlab{b}})}\BibitemShut
  {NoStop}%
\bibitem [{\citenamefont {Obermayer}\ \emph {et~al.}(2009)\citenamefont
  {Obermayer}, \citenamefont {M{\"o}bius}, \citenamefont {Hallatschek},
  \citenamefont {Frey},\ and\ \citenamefont {Kroy}}]{Obermayer:2009}%
  \BibitemOpen
  \bibfield  {author} {\bibinfo {author} {\bibfnamefont {B.}~\bibnamefont
  {Obermayer}}, \bibinfo {author} {\bibfnamefont {W.}~\bibnamefont
  {M{\"o}bius}}, \bibinfo {author} {\bibfnamefont {O.}~\bibnamefont
  {Hallatschek}}, \bibinfo {author} {\bibfnamefont {E.}~\bibnamefont {Frey}}, \
  and\ \bibinfo {author} {\bibfnamefont {K.}~\bibnamefont {Kroy}},\ }\href
  {\doibase 10.1103/PhysRevE.79.021804} {\bibfield  {journal} {\bibinfo
  {journal} {Physical Review E}\ }\textbf {\bibinfo {volume} {79}},\ \bibinfo
  {pages} {021804} (\bibinfo {year} {2009})}\BibitemShut {NoStop}%
\bibitem [{\citenamefont {Gardiner}(2009)}]{Gardiner:2009}%
  \BibitemOpen
  \bibfield  {author} {\bibinfo {author} {\bibfnamefont {C.}~\bibnamefont
  {Gardiner}},\ }\href {http://books.google.de/books?id=otg3PQAACAAJ} {\emph
  {\bibinfo {title} {Stochastic Methods: A Handbook for the Natural and Social
  Sciences}}},\ Springer Series in Synergetics\ (\bibinfo  {publisher}
  {Springer Berlin Heidelberg},\ \bibinfo {year} {2009})\BibitemShut {NoStop}%
\end{thebibliography}
%

\end{document}